\documentclass{article}

\usepackage[preprint]{neurips_2026}

\usepackage[utf8]{inputenc}
\usepackage[T1]{fontenc}
\usepackage{microtype}
\usepackage{inconsolata}

\usepackage{hyperref}
\usepackage{url}
\usepackage{booktabs}
\usepackage{amsfonts}
\usepackage{nicefrac}

\PassOptionsToPackage{dvipsnames,table}{xcolor}
\usepackage{xcolor}

\usepackage{amsmath}

\usepackage{amsmath,amsfonts,bm}

\def\eqref#1{equation~\ref{#1}}

\def\1{\bm{1}}

\DeclareMathAlphabet{\mathsfit}{\encodingdefault}{\sfdefault}{m}{sl}
\SetMathAlphabet{\mathsfit}{bold}{\encodingdefault}{\sfdefault}{bx}{n}

\def\gE{{\mathcal{E}}}

\def\gG{{\mathcal{G}}}

\def\gM{{\mathcal{M}}}

\def\gU{{\mathcal{U}}}

\DeclareMathOperator*{\argmin}{arg\,min}

\makeatletter
\DeclareRobustCommand\onedot{\futurelet\@let@token\@onedot}
\def\@onedot{\ifx\@let@token.\else.\null\fi\xspace}

\def\eg{\emph{e.g}\onedot} 
\def\ie{\emph{i.e}\onedot}

\makeatother
\usepackage{pifont}
\usepackage{xspace}

\newcommand{\pdb}{\textsc{PDB}\xspace}
\newcommand{\pdbsingle}{\textsc{PDB-Single-Full}\xspace}
\newcommand{\pdbsh}{\textsc{PDB-Single}\xspace}
\newcommand{\pdbmulti}{\textsc{PDB-Wild}\xspace}
\newcommand{\pdbhard}{\pdbsh}
\newcommand{\pdbfull}{\pdbsingle}
\newcommand{\pdblong}{\textsc{Precise Debugging Benchmarking}\xspace}
\newcommand{\bigcode}{BigCodeBench\xspace}
\newcommand{\livecode}{LiveCodeBench\xspace}
\newcommand{\swebench}{SWE-smith\xspace}

\newcommand{\apply}{\operatorname{apply}\xspace}
\newcommand{\map}{\operatorname{map}\xspace}
\newcommand{\ess}{\operatorname{essential}\xspace}

\usepackage{booktabs}

\usepackage{graphicx}
\usepackage{multirow}
\usepackage{array}
\usepackage{subcaption}
\usepackage{makecell}

\usepackage{enumitem}
\usepackage{titletoc}

\usepackage{listings}
\usepackage{tcolorbox}

\usepackage[ruled,vlined]{algorithm2e}

\usepackage{standalone}
\usepackage{tikz}
\usepackage{twemojis}
\usepackage{wrapfig}

\makeatletter
\newcommand\blfootnote[1]{%
  \begingroup
  \renewcommand\thefootnote{}%
  \footnotetext{#1}%
  \endgroup
}
\makeatother
\title{Precise Debugging Benchmark:\\ Is Your Model Debugging or Regenerating?}
\author{%
  Miaosen Chai$^{*\,\spadesuit}$ \quad
  Wang Bill Zhu$^{*\,\spadesuit}$ \quad
  Shangshang Wang$^{\spadesuit}$ \quad
  Yejia Liu$^{\S}$ \\
  \textbf{Song Bian}$^{\heartsuit}$ \quad
  \textbf{Honghua Dong}$^{\diamondsuit}$ \quad
  \textbf{Willie Neiswanger}$^{\spadesuit}$ \quad
  \textbf{Robin Jia}$^{\spadesuit}$ \\[0.25em]
  $^{\spadesuit}$University of Southern California \quad
  $^{\S}$Microsoft \\
  $^{\heartsuit}$University of Wisconsin--Madison \quad
  $^{\diamondsuit}$University of Toronto \\[0.5em]
  {
  \texttwemoji{chart increasing}~\href{https://huggingface.co/datasets/Precise-Debugging-Benchmarking/PDB-Single-Hard}{\textbf{Dataset}} \quad
  \texttwemoji{globe with meridians}~\href{https://precise-debugging-benchmark.github.io/}{\textbf{Webpage}} \quad
  \raisebox{-0.1em}{\includegraphics[width=1em]{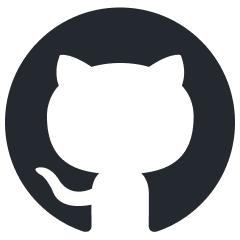}}~\href{https://github.com/Bill1235813/PDB}{\textbf{Code}} \quad
  \texttwemoji{trophy}~\href{https://precise-debugging-benchmark.github.io/leaderboard.html}{\textbf{Leaderboard}}%
  }%
}

\begin{document}

\maketitle
\blfootnote{$^{*}$\,Equal contribution.}
\blfootnote{$^{*}$\,Corresponding authors: \texttt{miaosenc@uchicago.edu}, \texttt{wangzhu@berkeley.edu}.}
\blfootnote{$^{\dagger}$\,Work done before joining Microsoft.}

\begin{abstract}
Unlike code completion, debugging requires localizing faults and applying targeted edits. 
We observe that frontier LLMs often hack unit tests by \textit{regenerating} correct but over-edited solutions during debugging.
To evaluate how far LLMs are from precise debugging, we introduce the \textbf{\pdblong (\pdb)}, a general, \textit{dataset-agnostic} framework that converts \textit{any} coding dataset into a debugging benchmark with precision-aware evaluation.
\pdb generates buggy programs by synthesizing verified atomic bugs and composing them into multi-bug programs.
We define two novel metrics, edit-level \textit{precision} and bug-level \textit{recall}, which measure how many necessary edits are made and how many bugs are resolved.
We release two evaluation benchmarks: \textbf{\pdbhard} on single-line bugs, and \textbf{\pdbmulti} on multi-line and repository-level bugs.
Experiments show that frontier models, such as GPT-5.1-Codex and DeepSeek-V3.2-Thinking, achieve unit-test pass rates above $76\%$ but exhibit precision below $45\%$, even when explicitly instructed to perform minimal debugging on single-line bugs. 
Iterative and agentic debugging strategies do not substantially improve precision or recall, highlighting the need to rethink post-training pipelines for coding models.


\end{abstract}


\section{Introduction}
\label{sec:intro}

\begin{wrapfigure}{r}{0.49\textwidth}
    \vspace{-1em}
    \centering
    \includegraphics[width=\linewidth]{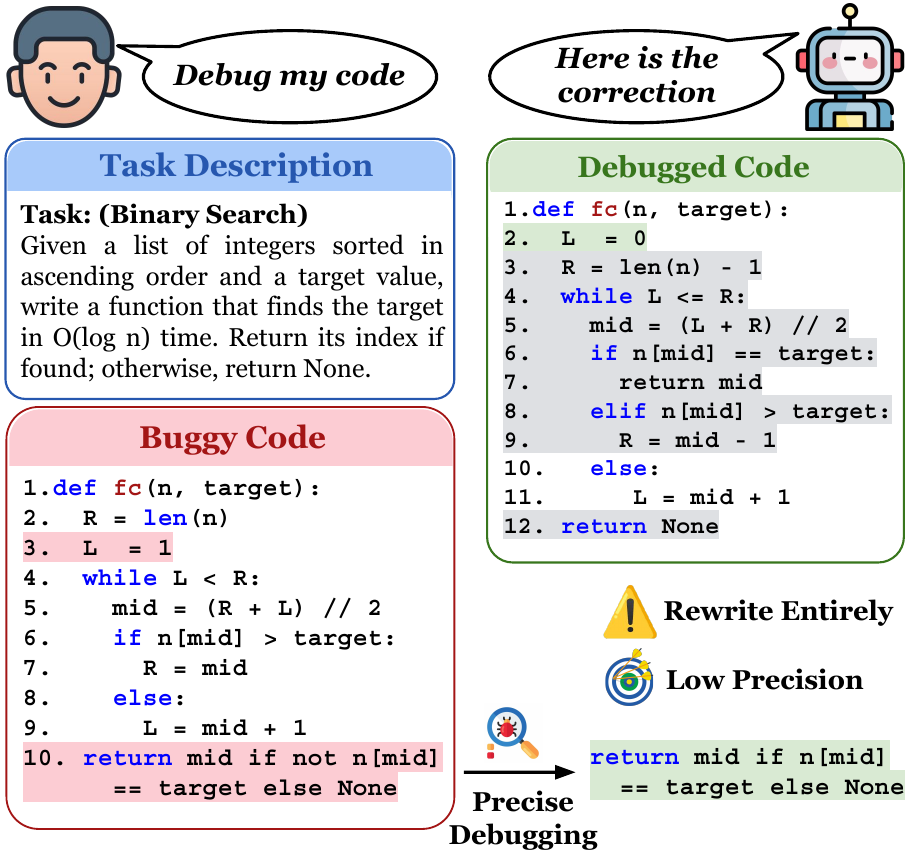}
    \caption{Real example from GPT-5.2 debugging a binary search program, where the model rewrites the entire solution. Green lines mark precise edits; gray lines highlight over-edits.}
    \label{fig:teaser}
    \vspace{-1em}
\end{wrapfigure}

Large Language Models (LLMs) have reshaped the programming landscape through their remarkable capabilities in code generation~\citep{chen2021evaluating, li2023starcoder}. 
From synthesizing complex algorithms from natural language prompts to translating entire codebases, modern LLMs excel at producing code from scratch. 
However, real-world software development is dominated not by generation but by debugging and maintenance~\citep{glass2002facts}.
When applied to debugging tasks, we observe that frontier LLMs often default to regeneration, \ie, rewriting large portions, or even the entirety, of a program when presented with buggy code (Figure~\ref{fig:teaser}). 
While often effective at passing tests, this brute-force strategy is poorly suited for realistic codebases, where large-scale rewrites are costly, risky, and difficult to review~\citep{sobania2023analysis}. 
In contrast, targeted debugging requires precise fault localization and minimal, intent-preserving edits. 
This raises a fundamental question: \textit{How far are LLMs from precise debugging, rather than merely reverting to their strength in code regeneration?}

Existing debugging benchmarks focus on unit-test only evaluation and fail to evaluate these capabilities.
Under such evaluation, models are rewarded equally for regenerating a full solution, hard-coding outputs, or performing a minimal targeted fix. Moreover, unit-test evaluation obscures incremental progress: a model that correctly repairs only one defect in a multi-bug program receives the same score as a model that fixes none. 
This misalignment with real-world debugging practice limits our ability to understand how LLMs reason about bugs and code edits. Also, existing debugging benchmarks are predominantly derived from publicly accessible debugging questions and answers collected from platforms such as Stack Overflow and GitHub issues~\citep{jimenez2023swe,yasunaga2021break,phan2021cotext}. They are vulnerable to contamination and require extensive manual effort to construct.

To address these gaps, we introduce the \pdblong (\pdb) framework, a general pipeline that rigorously evaluates LLM debugging behavior independently of code generation. \pdb provides a dataset-agnostic plug-and-play framework that converts existing coding datasets into debugging benchmarks through two steps: (1) synthesizing verified atomic bugs to produce ground-truth edit scripts, and (2) composing these bugs into multi-bug programs while preserving bug independence (\ie, avoiding compounding interactions). Beyond binary test outcomes, \pdb evaluates model patches using novel edit-level \textit{precision} and bug-level \textit{recall}, explicitly rewarding targeted fixes and penalizing unnecessary modifications. 

We constructed a 5,751-example \pdbhard and a 484-example \pdbmulti evaluation benchmarks with the \pdb framework, from \bigcode~\citep{zhuo2024bigcodebench}, \livecode~\citep{jain2024livecodebench}, and \swebench~\citep{yang2025swe}.
Experiments on \pdbhard reveal behaviors that unit tests fail to capture.
First, frontier models exhibit strikingly \textbf{different rankings under edit-level evaluation}. 
Models such as GPT-5.1-Codex \cite{gpt-5.1-codex} and DeepSeek-V3.2-Thinking \cite{liu2025deepseek} achieve high unit-test pass rates ($>$$76\%$) but low edit precision ($\leq$$45\%$), while Qwen3-Coder-480B \cite{qwen3-coder} attains comparatively lower unit-test pass rates ($70\%$) yet substantially higher precision ($66\%$).
This ranking inversion \textbf{persists} on the multi-line and repository-level \pdbmulti{} benchmark, including when newer models such as GPT-5.5 and Claude-Opus-4.7 are evaluated on bugs generated by the models themselves.
Additionally, we show that though iterative and agentic debugging strategies can improve unit-test performance, they do not meaningfully improve precision or recall.
Our findings demonstrate the necessity of \pdb for revealing true debugging capabilities beyond surface-level correctness, and highlight a fundamental limitation in current post-training pipelines for coding LLMs.

\section{Precise Debugging Setup}
\label{sec:setup}

\begin{figure*}[t]
    \centering
    \includegraphics[width=\linewidth]{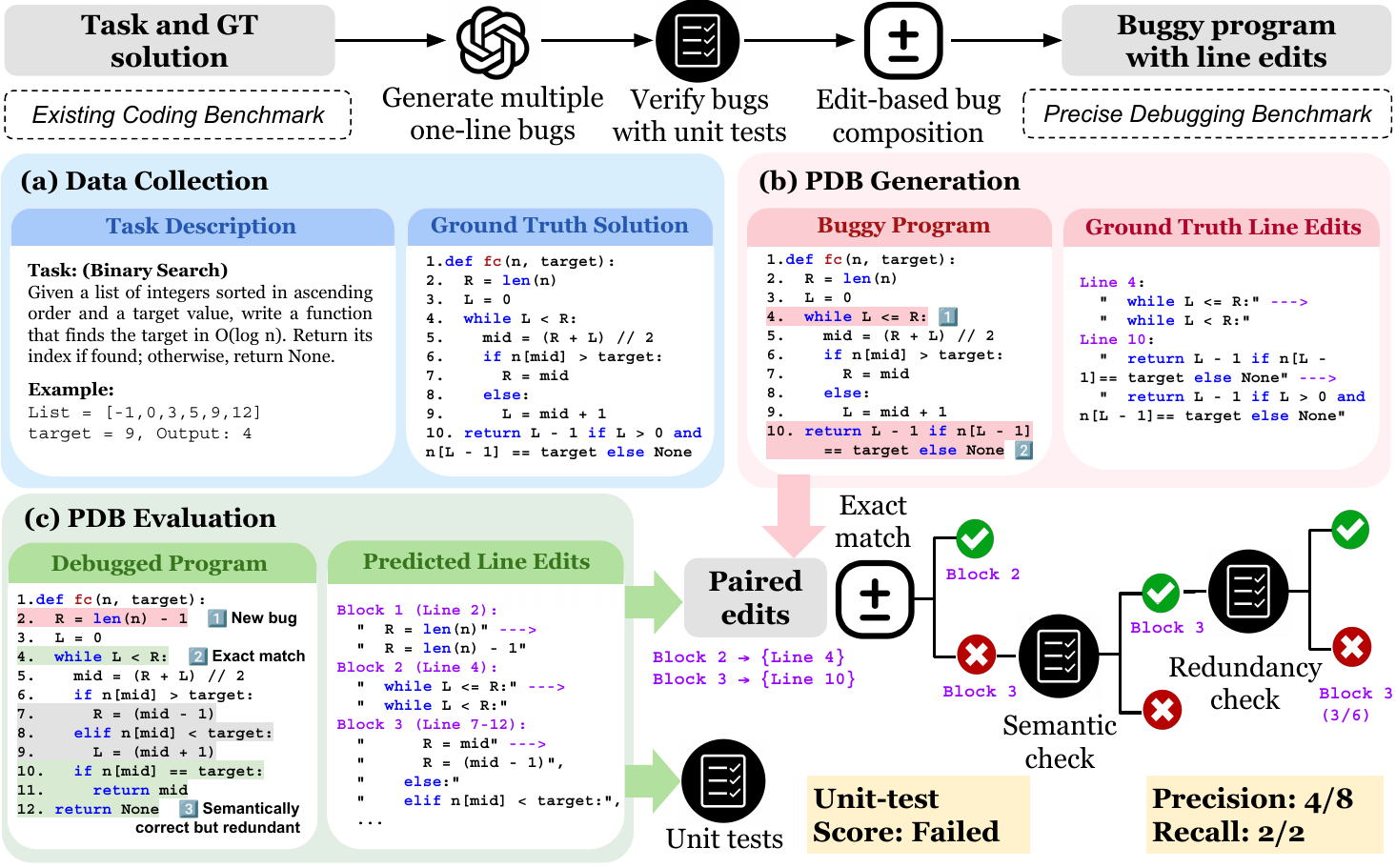}
    \caption{\pdb pipeline. Generation: LLMs first synthesize and verify single-line bugs from existing coding datasets, which are then composed into multi-bug programs. Evaluation: Automated debugging systems are evaluated on these programs using both unit-test accuracy and edit-level precision and bug-level recall.}
    \label{fig:pipeline}
\end{figure*}

We begin by formally defining the components of the automated debugging task. An automated debugging system $\mathcal{M}$, takes the initial buggy program $C_{\text{b}}$ and a natural language task description $x$ as input, and returns the predicted program revision $\hat{C}=\mathcal{M}(C_{\text{b}}, x)$.
The conventional debugging pipeline evaluates the system's final output purely on its functional correctness, using a binary evaluation function $F_{\gU}(C) \rightarrow \{0, 1\}$, where $\gU = \{u_1, u_2, \dots, u_n\}$ is a suite of designed unit tests. 
The evaluation function $F_{\gU}$ returns 1 if the program $C$ passes all tests in $\gU$, and 0 otherwise.

While straightforward, this method cannot penalize unnecessary edits or wholesale rewrites when the final program merely passes the test, nor can it distinguish between partially correct solutions and entirely incorrect ones.
The precise debugging setup shifts the evaluation from program-level to the specific set of edits proposed by the model.

\noindent\textbf{Minimal corrections.} We denote a line-edit on line $l$ as $e_l$, and a set of line-edits as $E$. 
For a buggy program $C_{\text{b}}$, we denote the set of \textit{minimal corrections} by $\gE_{C_\text{b}} = \argmin_{E} |E| \ \text{s.t.} \ F_{\gU}(\apply(E, C_{\text{b}})) = 1,$
where the $\apply$ function applies line-edits on $C_{\text{b}}$.
Similarly, we can apply reverse edits $\bar{E}$ on ground-truth program $C_{\text{gt}}$ to derive buggy program $C_{\text{b}}$.

\noindent\textbf{Atomicity.}
We define the bug in a buggy program $C_{\text{b}}$ as \emph{atomic} when a minimal correction consists of edits on a contiguous sequence of lines. Formally, $\exists E \in \gE_{C_{\text{b}}}$ such that $E = \{ e_i, e_{i+1}, \dots, e_{i+n} \}.$

\noindent\textbf{Independence.}
Intuitively, independence means that fixing one bug neither introduces nor removes edits required to fix the other.
For two edit sets $E_1 \in \gE_{C_{\text{b1}}}$ and $E_2 \in \gE_{C_{\text{b2}}}$ corresponding to the same ground-truth program $C_{\text{gt}}$, we can construct a composed buggy program $C_{\text{b3}} = \apply(\bar{E}_1 \cup \bar{E}_2, C_{\text{gt}}).$
If the set of minimal corrections is the pairwise union of corrections from $\gE_{C_{\text{b1}}}$ and $\gE_{C_{\text{b2}}}$,
we consider bugs in $C_{\text{b1}}$ and $C_{\text{b2}}$ to be \emph{independent}.

\noindent\textbf{Semantic correctness.}
Consider a buggy program $C_{\text{b}}$ containing $k$ atomic and independent bugs, and a revision $\hat{C} = \apply(\hat{E}, C_{\text{b}})$, where $\hat{E}$ is the predicted edits. 
We define \emph{bug-level semantic correctness} as follows.

Let $E_{\text{gt}} \in \gE_{C_{\text{b}}}$ be the set of ground-truth edits, which can be decomposed as
$E_{\text{gt}} = E_1 \cup E_2 \cup \dots \cup E_k$, where $E_1, \dotsc, E_k$ are contiguous and non-overlapping. 
We employ a function, denoted as $\map$, which pairs each $E_i$ with the closest edits in $\hat{E}$. 
For each bug $i$, we construct a pseudo-revision
$\hat{C}_i = \apply((E_{\text{gt}} \setminus E_i) \cup \map(E_i),\, C_{\text{b}})$, which replaces the ground-truth edits $E_i$ with the predicted edits $\map(E_i)$.  
We define a candidate $\hat{C}_i$ as \textit{semantically correct} for bug $i$ if $F_{\gU}(\hat{C}_i) = 1$. Figure~\ref{fig:pipeline} (Block 3) is such an example. 
Based on this, we define precision and recall as:
\begin{align}
   \text{precision} & = \frac{1}{|\hat{E}|}\sum_{i=1}^{k} F_{\gU}(\hat{C}_i)\cdot|E_i|,
   \label{eqn:precision} \\
   \text{recall} & = \frac{1}{k}\sum_{i=1}^{k} F_{\gU}(\hat{C}_i).
   \label{eqn:recall}
\end{align}
We note that precision functions as an edit-level metric by averaging over the edits $|\hat{E}|$, while recall is a bug-level metric averaged over the $k$ bugs.

\noindent\textbf{$\epsilon$-relaxed essential edits.}
Since our objective is to discourage solution \emph{regeneration} rather than to enforce \emph{strictly minimal} edits, we relax the precision metric in Eq.~(\ref{eqn:precision}) by introducing a tolerance parameter $\epsilon$, which allows up to $|E_i|+\epsilon$ edited lines for each bug $i$.

Moreover, even when a candidate revision $\hat{C}_i$ is semantically correct for bug $i$, the predicted edits $\map(E_i)$ may still contain regeneration, as illustrated in Figure~\ref{fig:pipeline} (Block 3). To remove such redundancy, we introduce a unit-test–based function $\ess_{\gU}$, which searches over subsets of $\map(E_i)$ to recover the minimal essential edits required to resolve bug $i$ while preserving semantic correctness. Formally, we define the \emph{$\epsilon$-relaxed essential edit size} for bug $i$ as
\begin{align*}
    (|\hat{E}_i|)_\epsilon = \min (|\ess_\gU(\map(E_i))|, |E_i| + \epsilon).
\end{align*}
Accordingly, the $\epsilon$-relaxed precision is defined as
\begin{align}
    \text{precision}_{\epsilon}
    =
    \frac{1}{|\hat{E}|}
    \sum_{i=1}^{k}
    F_{\gU}(\hat{C}_i)\cdot (|\hat{E}_i|)_{\epsilon}.
    \label{eqn:precision_relax}
\end{align}
We provide full details of the $\map$ and $\ess_{\gU}$ procedures in Appendix~\ref{appsec:algo}.
\section{Generation and Evaluation Pipeline}
\label{sec:pipeline}

As illustrated in Figure~\ref{fig:pipeline}, \pdb consists of two stages: \emph{generation} and \emph{evaluation}. 
During the \pdb generation stage, we first use LLMs to synthesize atomic bugs from existing coding datasets. 
After verifying buggy programs with unit tests, we record the corresponding edit sets and compose them to construct multi-bug programs.
During the \pdb evaluation stage, we prompt an automated debugging system $\gM$ to revise the buggy programs, and evaluate its performance using both traditional unit-test accuracy and our proposed edit-level precision and bug-level recall metrics.
All prompt templates are provided in Appendix~\ref{appsec:prompt}.

\subsection{\pdb generation}
Starting from an existing coding benchmark, 
for each task description $x$ and ground-truth program $C_{\text{gt}}$, we generate buggy programs across five Orthogonal Defect Classification (ODC; \citealp{chillarege1992orthogonal}) categories: \textit{Assignment}, \textit{Checking}, \textit{Algorithm}, \textit{Build/Package/Merge}, and \textit{Timing/Serialization}. Each category further contains several subcategories, listed in Table~\ref{tab:odc_taxonomy}.

\noindent\textbf{Atomic bug generation.}
Single-line bug can ensure atomicity, as the minimal correction satisfies $|E| = 1$.
We consider three types of line-level operations: \textit{insertion}, \textit{deletion}, and \textit{substitution}.
We first apply a rule-based filter to identify lines that are not safely deletable (\eg, causing indentation errors) or not editable (\eg, function headers).
To promote diversity, we then randomly select (i) one operation type, (ii) one bug category, and (iii) a subset of editable lines compatible with the chosen operation.
An LLM from a generator pool is prompted to modify one of the selected lines to produce a single-line buggy program.
We repeat this process $m_1$ times per $(x, C_{\text{gt}})$ pair and retain only programs that fail unit tests, ensuring the validity of injected bugs.

\noindent\textbf{Multiline bug generation.}
To extend the pipeline to multi-line bugs, we apply the same editing procedure to contiguous blocks of code.
Specifically, we randomly select (i) a block size $B \in [2, B_{\max}]$, (ii) one primary and two auxiliary bug categories, and (iii) a valid range from which to sample a contiguous block of lines.
Because multi-line edits --- even within a contiguous block --- do not inherently guarantee atomicity, we explicitly filter out violations.
We implement an \emph{atomicity filter} by enumerating all partial fixes that revert a strict subset of the modified lines back to $C_{\text{gt}}$, and retain a bug instance only if \emph{all} such partial fixes still fail unit tests.
This procedure removes non-atomic cases where fixing a subset of edits is sufficient to pass the tests, which would otherwise inflate both precision and recall.

\noindent\textbf{Bug composition.}
To create more challenging debugging scenarios, we compose multiple atomic bugs into a single program.
For each $(x, C_{\text{gt}})$ pair and a target bug count $k$, we randomly sample $k$ distinct block edits from the generated bugs.
To encourage independence between bugs, we enforce a \emph{stride} constraint, requiring any two selected edits to be at least $s$ lines apart.
For each bug count $k \in \{2, \ldots, k_{\max}\}$, we repeat this process $m_2$ times per $(x, C_{\text{gt}})$ pair and record all composed multi-bug programs that satisfy the constraint.

\noindent\textbf{Subsampling.}
To avoid over-representation of $(x, C_{\text{gt}})$ pairs with many generations, we subsample the data by randomly selecting at most $m_3$ buggy programs per bug count per $(x, C_{\text{gt}})$ pair.

\subsection{\pdb evaluation}
During evaluation, debugging systems, either single-pass LLMs or LLM-based agents, are instructed to debug a buggy program $C_{\text{b}}$, given the task description $x$ and, optionally, access to unit tests $\gU$ and unit-test error feedback.

We use the precision, recall equations as in Eq.~(\ref{eqn:recall}, \ref{eqn:precision_relax}), and report the unit-test score at Pass@1 \citep{kulal2019spoc}.
All metrics are reported as averages over all examples in an evaluation set.

\subsection{\pdb test-suite}
PDB separates two properties that are sometimes conflated.  First,
\emph{instance detectability} asks whether the inherited tests expose the
specific injected bug.  PDB enforces this property per instance: a candidate
is retained only if the complete injected program fails, and a multi-line
candidate is atomic only if every strict partial repair also fails.  A suite
that cannot detect a proposed bug therefore cannot admit that bug to the
benchmark.  Second, \emph{scoring adequacy} asks whether every test-passing
model patch is functionally correct. When instantiating PDB on a new source dataset, test adequacy should be treated
as an upstream requirement. 

Although no finite execution-based benchmark can guarantee this property in general, we empirically estimate the false-positive rate of our released dataset to be 1.85\% due to the upstream test-suite. Here, a false positive refers to a model-generated patch that passes the inherited tests despite being functionally incorrect. More details in \ref{appsec:unit-test}. 

We also add an optional test-adequacy gate to the released pipeline's preprocessing stage. \emph{Coverage gate}: before bug injection, run the suite on the ground-truth solution under a line tracer and require every editable line to be executed by at least one test; lines never executed cannot host a detectable bug and are excluded from injection. \emph{Test augmentation}: for tasks that fail the coverage gate, generate additional tests with an LLM and keep only those that pass on the ground-truth solution in repeated runs, optionally as property-based tests; the augmented suite then feeds the standard PDB validation and scoring path. 

\section{Evaluation Sets}
\label{sec:eval_sets}

\begin{figure*}[t]
    \centering
    \begin{subfigure}[hb]{0.43\linewidth}
        \centering
        \includegraphics[width=\linewidth]{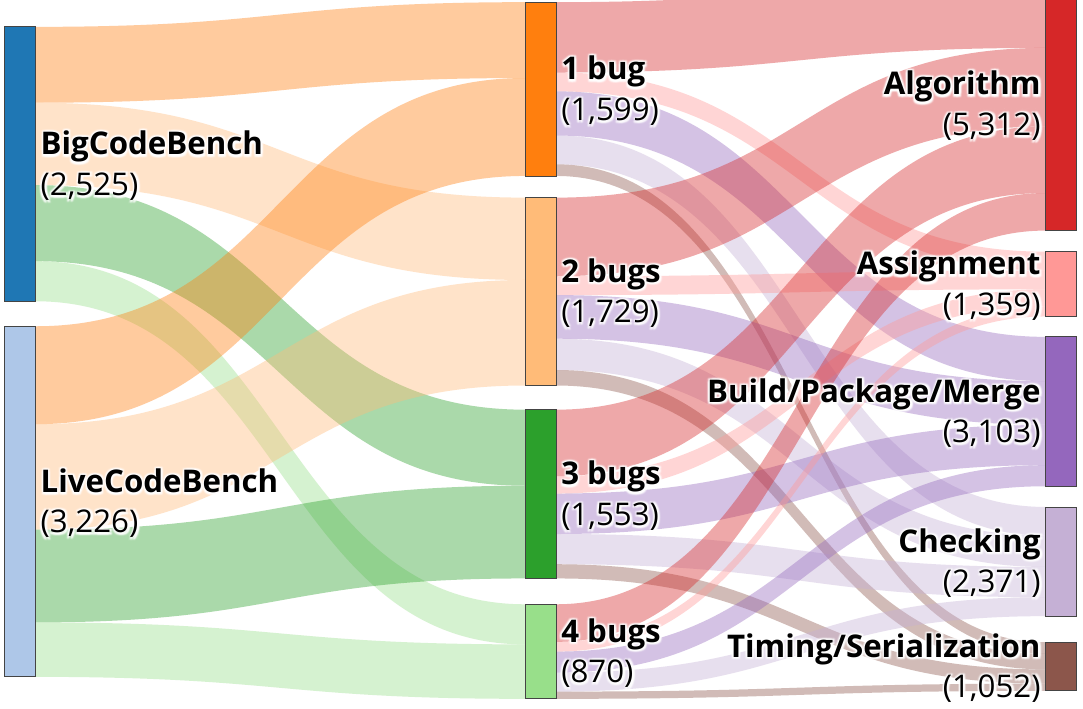}
        \caption{}
        \label{fig:sankey_data_distribution}
    \end{subfigure}
    \hfill
    \begin{subfigure}[hb]{0.55\linewidth}
        \centering
\tabcolsep 5pt
\resizebox{\linewidth}{!}{
\begin{tabular}{@{}lrrr@{}}
\toprule
\bf Model & \bf Precision  & \bf Recall & \bf Unit (\%) \\
\midrule
Claude-Sonnet-4.5 & \color[rgb]{0.000,0.000,1.000}\textbf{71.8$\pm$0.9} & \color[rgb]{0.146,0.000,0.854}81.4$\pm$0.8 & \color[rgb]{0.200,0.000,0.800}75.7$\pm$1.1 \\
Gemini-2.5-Pro & \color[rgb]{0.005,0.000,0.995}71.4$\pm$0.9 & \color[rgb]{0.000,0.000,1.000}\textbf{83.5$\pm$0.7} & \color[rgb]{0.068,0.000,0.932}78.1$\pm$1.0 \\
Qwen3-Coder-480B & \color[rgb]{0.171,0.000,0.829}65.8$\pm$0.9 & \color[rgb]{0.368,0.000,0.632}77.2$\pm$0.9 & \color[rgb]{0.456,0.000,0.544}70.3$\pm$1.2 \\
Kimi-K2-Instruct & \color[rgb]{0.455,0.000,0.545}56.6$\pm$1.0 & \color[rgb]{0.602,0.000,0.398}72.7$\pm$0.9 & \color[rgb]{0.704,0.000,0.296}64.8$\pm$1.2 \\
Grok-Code-Fast & \color[rgb]{0.539,0.000,0.461}54.6$\pm$1.0 & \color[rgb]{1.000,0.000,0.000}66.5$\pm$1.0 & \color[rgb]{1.000,0.000,0.000}58.3$\pm$1.3 \\
Kimi-K2-Thinking & \color[rgb]{0.637,0.000,0.363}51.7$\pm$0.9 & \color[rgb]{0.449,0.000,0.551}75.6$\pm$0.9 & \color[rgb]{0.245,0.000,0.755}74.0$\pm$1.1 \\
DeepSeek-V3.2 & \color[rgb]{0.756,0.000,0.244}48.4$\pm$1.0 & \color[rgb]{0.823,0.000,0.177}70.0$\pm$1.0 & \color[rgb]{0.397,0.000,0.603}71.4$\pm$1.2 \\
DeepSeek-V3.2-Thinking & \color[rgb]{0.852,0.000,0.148}45.0$\pm$0.9 & \color[rgb]{0.760,0.000,0.240}71.2$\pm$1.0 & \color[rgb]{0.000,0.000,1.000}\textbf{79.0$\pm$1.0} \\
GPT-5.1-Codex & \color[rgb]{1.000,0.000,0.000}39.7$\pm$0.8 & \color[rgb]{0.711,0.000,0.289}71.7$\pm$0.9 & \color[rgb]{0.138,0.000,0.862}76.1$\pm$1.1 \\
\bottomrule
\end{tabular}
}
\caption{}
\label{tab:model_performance}
    \end{subfigure}
    \caption{(a) Data distribution of \pdbhard. (b). Precision, recall, and unit score on the \pdbhard set. Blue indicates better performance, while red indicates worse.}
    \label{fig:bug-count-breakdown}
\end{figure*}

Using the \pdb{} generation pipeline, we release two evaluation sets: \pdbsh{} targets single-line bugs, while \pdbmulti{} extends the pipeline to contiguous multi-line bug blocks. 
We source tasks from three existing coding benchmarks: \bigcode~\citep{zhuo2024bigcodebench} for API-call-heavy tasks, \livecode~\citep{jain2024livecodebench} for competitive programming, and \swebench~\citep{yang2025swe} for repository-level software engineering. Our bug-generation pool consists of four frontier LLMs: GPT-5.1-Codex~\citep{gpt-5.1-codex}, Claude-Sonnet-4.5~\citep{claude-sonnet-4.5}, Gemini-2.5-Pro~\citep{comanici2025gemini}, and Claude-Opus-4.7 for \swebench{}-specific generation.

\noindent\textbf{\pdbsh.}
For each ground-truth task we generate $m_1 = 20$ single-line bugs, compose up to $m_2 = 100$ multi-bug variants with at most $k_{\max} = 4$ independent bugs per program, and subsample $m_3 = 5$ buggy programs per bug count per task. A stride of $s=3$ lines is enforced between composed blocks. This yields the initial \pdbsingle set of 7,589 examples (see Appendix~\ref{appsec:pdbfull} for details).
We evaluate \pdbsingle on 9 models, including thinking models: GPT-5.1-Codex, Claude-Sonnet-4.5, Gemini-2.5-Pro, Grok-Code-Fast \citep{grok-code-fast-1}, DeepSeek-V3.2-Thinking, and Kimi-K2-Thinking~\citep{team2025kimi}; and non-thinking models: Qwen3-Coder-480B, DeepSeek-V3.2, and Kimi-K2-Instruct~\citep{team2025kimi}. All models are \textit{prompted to produce minimal code edits}. We use a maximum output length of 32,000 tokens for thinking models and 8,000 tokens for non-thinking models, with a temperature of 1.0 throughout.
We then apply model-based filtering to identify easy examples. 
We use a tolerance $\epsilon=2$ for precision evaluation with Eq.~(\ref{eqn:precision_relax})
An example is labeled easy if it achieves perfect precision, recall, and unit-test score for at least 7 out of the 9 evaluated models. Applying this criterion removes 1,838 examples, resulting in the final \pdbhard benchmark of 5,751 challenging examples.


\noindent\textbf{\pdbmulti.}
For multi-line bugs, we select programs from \bigcode{} and \livecode{} with more than 35 lines and assign each generator a disjoint subset.
We set the maximum block size to $B_{\max}=4$, use stride $s=5$ with the same $m_1,m_2,m_3$, and compose up to $k_{\max}=3$ blocks per program, yielding 256 examples.
To cover realistic software-engineering scenarios, we further sample 10 repositories from 128 public \swebench{} repositories~\citep{yang2024swe,yang2025swe}; for each repository, we extract up to ten non-test source files and set $B_{\max}=30$, $k_{\max}=3$. Validated bugs survive in 32 files (110--1{,}222 lines, median 378) from 6 of the repositories, yielding 228 examples.
Together, \pdbmulti{} contains 484 examples.
Because multi-line blocks cannot strictly guarantee atomicity (for the up-to-30-line \swebench{} blocks we check single-line partial repairs only), we use a tolerance of $\epsilon=1$.

\section{Experiment Results}
\label{sec:results}

By evaluating and analyzing both LLMs and LLM-based agents on \pdbhard and \pdbmulti, we show that current systems remain far from achieving precise, edit-aware debugging.

\begin{figure*}[t]
    \centering
    \begin{subfigure}[t]{0.49\linewidth}
        \centering
        \includegraphics[width=\linewidth]{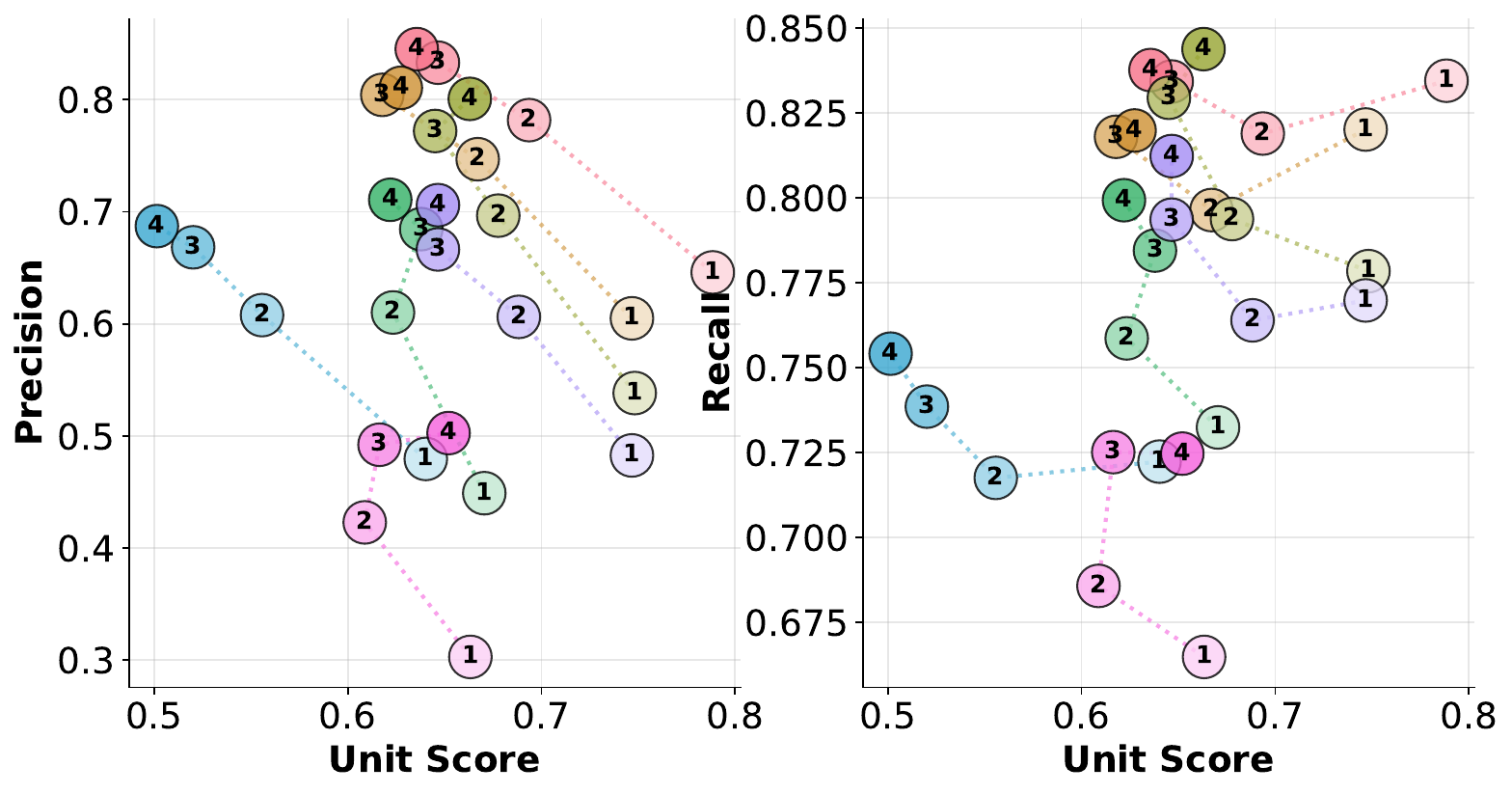}
    \end{subfigure}
    \hfill
    \begin{subfigure}[t]{0.49\linewidth}
        \centering
        \includegraphics[width=\linewidth]{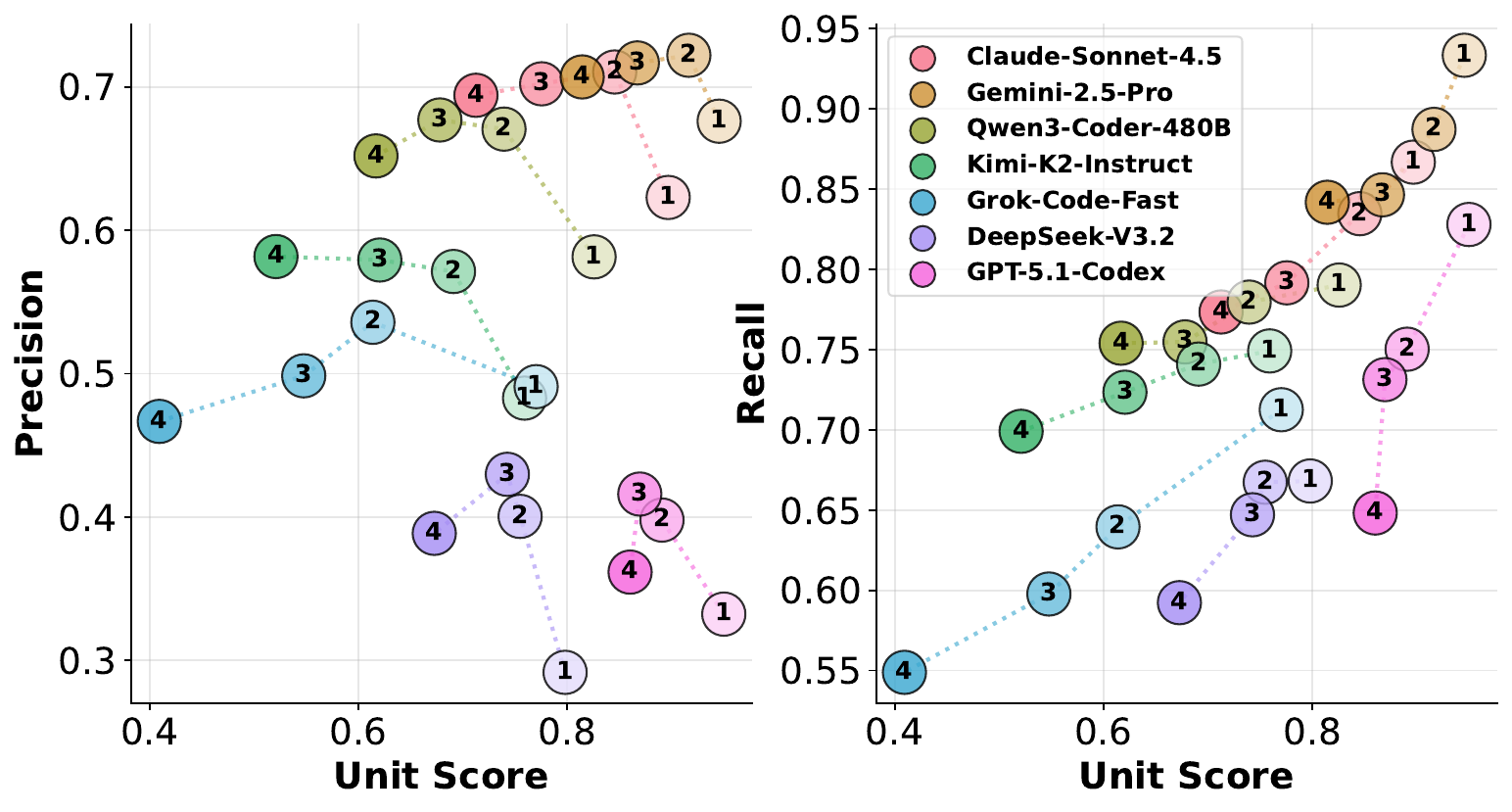}
    \end{subfigure}
    \caption{Correlation between precision, recall, and unit-test score across bug counts. Results are shown on subsets of \pdbhard from \bigcode (left) and \livecode (right), with bug counts indicated by numbers. As the number of bugs increases, precision generally exhibits a negative correlation with unit-test score, while recall displays dataset-dependent behavior.}
    \label{fig:bug-count-breakdown}
\end{figure*}

\subsection{\pdbhard overview}

\noindent\textbf{Divergence in model debugging behaviors.}
Even among top-performing models such as Claude-Sonnet-4.5 and Gemini-2.5-Pro, performance can only be characterized as relatively precise and faithful: no frontier model exceeds $72\%$ precision, even when explicitly instructed to perform minimal debugging.
Table~\ref{tab:model_performance} shows that unlike unit-test pass rates, edit-level precision and bug-level recall reveals four different types of model debugging behaviors:
\begin{itemize}[leftmargin=*, itemsep=0pt, topsep=0pt, nosep]
    \item \textit{Pass with precision}: Claude-Sonnet-4.5 and Gemini-2.5-Pro debug correctly ($>$$75\%$) with the highest precision ($>$$71\%$) and recall ($>$$81\%$).
    \item \textit{Weak but precise}: Qwen3-Coder-480B, though only achieves $70\%$ unit score, has moderately high precision ($66\%$) and recall ($77\%$).
    \item \textit{Weak, imprecise, but identifying}: Kimi-K2-Instruct, Kimi-K2-Thinking, and Grok-Code-Fast reliably identify buggy regions but struggle to produce correct and precise fixes, with precision below $57\%$.
    \item \textit{Pass-oriented}: DeepSeek-V3.2, DeepSeek-V3.2-Thinking, and GPT-5.1-Codex exhibit substantially lower precision ($\leq\!48\%$), with recall below unit test scores, indicating a regeneration-heavy behavior that relies on broad rewrites.
\end{itemize}
These results highlight the necessity of edit-level evaluation for distinguishing targeted debugging behavior from superficial pass-driven regeneration.

\noindent\textbf{Negative correlation between unit score and precision.}
We further analyze model performance as the number of injected bugs increases ($k \in \{1,2,3,4\}$), corresponding to increasing problem complexity. 
As shown in Figure~\ref{fig:bug-count-breakdown}, unit-test scores consistently decrease across all models as the number of bugs increases. At the same time, we observe an inverse trend for edit-level precision. 
Because models tend to over-edit, increasing the number of bugs raises the likelihood that a model modifies at least one necessary line, but also increases the amount of unnecessary edits, leading to lower precision overall. 
This trend is further supported by our analysis in Appendix~\ref{appsec:exps}, which shows that precision degrades as buggy code length increases.

\begin{wrapfigure}{r}{0.4\linewidth}
    \vspace{-1.5em}
    \centering
    \includegraphics[width=\linewidth]{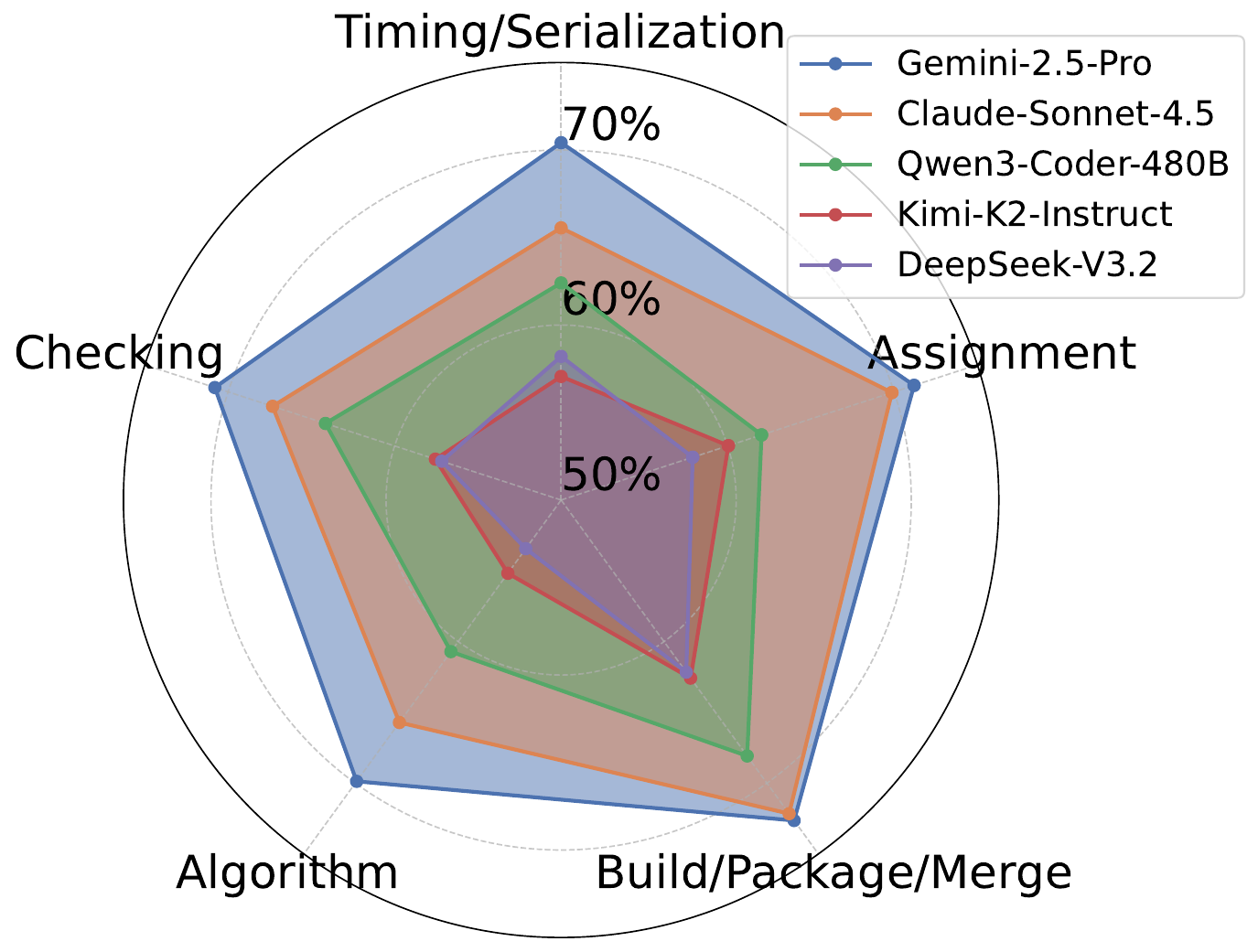}
    \caption{Bug-level recall by ODC category for five models spanning the four behavior groups above. The radial axis starts at $50\%$.}
    \label{fig:bug_type_breakdown}
    \vspace{-1em}
\end{wrapfigure}

In contrast, recall primarily reflects debugging difficulty per bug, as it measures the fraction of bugs successfully addressed. 
Consistent with this interpretation, recall exhibits dataset-dependent behavior. 
On the API-focused \bigcode benchmark, where the difficulty of fixing individual bugs remains relatively stable, recall varies by less than $5\%$ across bug counts from 1 to 4. 
On the algorithm-focused \livecode benchmark, where debugging difficulty increases with the number of injected bugs, recall shows a clear positive correlation with unit-test scores.

\noindent\textbf{Recall across defect categories.}
Because bugs are generated per ODC category (\S\ref{sec:pipeline}), \pdb also supports category-level analysis.
As shown in Figure~\ref{fig:bug_type_breakdown}, with the exception of Gemini-2.5-Pro, whose recall is relatively uniform across categories, models achieve their highest recall on \textit{Build/Package/Merge} defects.
We hypothesize that this advantage arises from the higher prevalence of such defects, \eg, invalid API calls and dependency-version conflicts, in model pretraining data, making them easier for models to recognize and repair.
Appendix~\ref{appsec:related} further relates these categories to classical mutation operators (Table~\ref{tab:mutation_odc}).

\begin{figure*}[t]
    \centering
    \includegraphics[width=\linewidth]{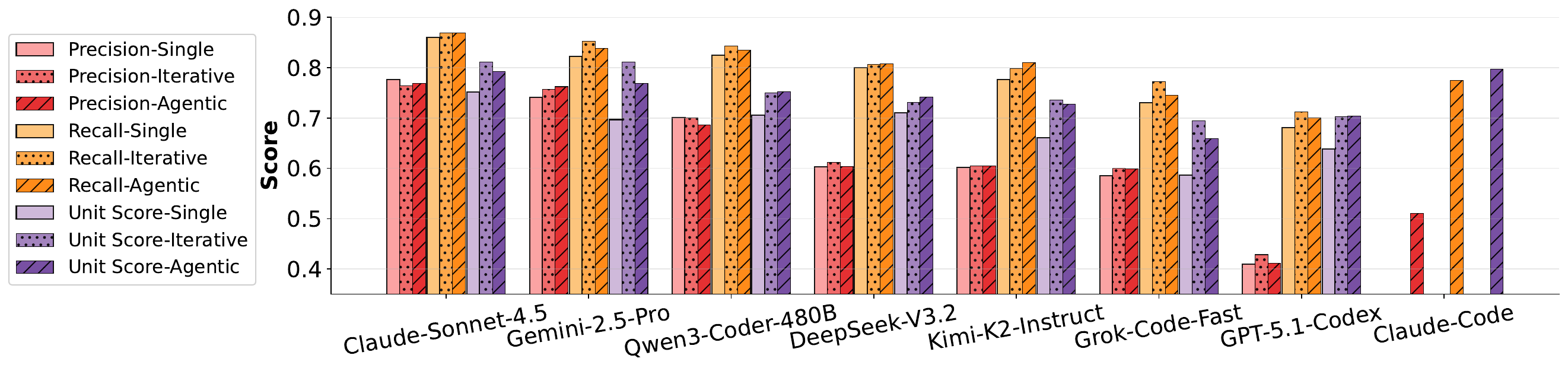}
    \caption{Both iterative and agentic setups on \pdbhard improve unit-test pass rates and recall over single-shot debugging, indicating higher functional success. However, edit-level precision does not improve and sometimes degrades. Notably, even Claude-Code with access to unit-test and execution feedback exhibits only $50\%$ precision.}
    \label{fig:ablation_multirun}
\end{figure*}

\subsection{Iterative and agentic debugging}
Next, we evaluate model behavior under \emph{iterative} and \emph{agentic} debugging settings. In \emph{iterative debugging}, models produce an initial single-shot solution and are then allowed up to three revision attempts per problem. 
The process terminates early if a revision passes all unit tests. Models have access to their previous failed outputs, approximating an interactive debugging workflow commonly used by human programmers. 
In the \emph{agentic} setting, models are likewise permitted up to three attempts, but additionally receive unit tests and execution error feedback at each step, resembling a simple agentic debugging pipeline with explicit external feedback.
We evaluate on 500 instances randomly sampled from \pdbhard, \bigcode sourced subset.

\noindent\textbf{Functional gains without precision improvements.}
As shown in Figure~\ref{fig:ablation_multirun}, both iterative and agentic settings consistently improve unit-test scores and recall, indicating a higher likelihood of eventually producing functionally correct programs and resolving a larger fraction of bugs. However, these gains do not translate into improved edit-level precision. In most cases, precision remains unchanged or degrades relative to single-shot debugging. This pattern suggests that iterative interaction primarily improves correctness by expanding the scope of code modifications, rather than by refining or localizing edits toward minimal repairs.

\noindent\textbf{Ineffective use of feedback in agentic debugging.}
Despite direct access to unit tests and execution feedback, most models fail to leverage this information to improve edit-level behavior in the agentic setting (Figure~\ref{fig:ablation_multirun}). In particular, agentic debugging often underperforms iterative debugging in precision, suggesting that additional feedback may exacerbate regeneration-oriented behaviors. Rather than supporting fault localization, test outcomes and error messages are frequently treated as coarse success signals that trigger further broad rewrites. These results indicate that access to feedback alone is insufficient to induce edit-aware debugging.

\noindent\textbf{Regeneration persists even in Claude-Code.}
We observe similar trends in Claude-Code, which achieves the highest precision among agentic methods but still attains only approximately $50\%$ precision. As shown in Figure~\ref{fig:ablation_multirun}, this result indicates that even more sophisticated, end-to-end agentic systems largely rely on regeneration rather than precise editing, reinforcing the conclusion that current debugging agents lack robust mechanisms for localized, minimal code repair.

\subsection{Multi-line bug extension}
\label{sec:pdb_multi_results}

\begin{wraptable}{r}{0.47\linewidth}
    \vspace{-2em}
    \centering
\centering
\tabcolsep 2pt
\caption{Performance on \pdbmulti with tolerance $\epsilon=1$. The same precision gap persists under multi-line and repo-level bugs.}
\label{tab:pdb_multi_results}
\resizebox{\linewidth}{!}{
\begin{tabular}{@{}lrrr@{}}
\toprule
\bf Model & \bf Precision & \bf Recall & \bf Unit (\%) \\
\midrule
Claude-Opus-4.7 & \color[rgb]{0.000,0.000,1.000}\textbf{77.8} & \color[rgb]{0.083,0.000,0.917}83.4 & \color[rgb]{0.223,0.000,0.777}75.8 \\
Gemini-3.1-Pro-Preview & \color[rgb]{0.002,0.000,0.998}77.8 & \color[rgb]{0.000,0.000,1.000}\textbf{85.6} & \color[rgb]{0.000,0.000,1.000}\textbf{85.1} \\
Claude-Sonnet-4.5 & \color[rgb]{0.224,0.000,0.776}68.7 & \color[rgb]{0.294,0.000,0.706}77.8 & \color[rgb]{0.381,0.000,0.619}69.2 \\
GPT-5.5 & \color[rgb]{0.415,0.000,0.585}60.9 & \color[rgb]{0.099,0.000,0.901}83.0 & \color[rgb]{0.020,0.000,0.980}84.3 \\
Gemini-2.5-Pro & \color[rgb]{0.480,0.000,0.520}58.3 & \color[rgb]{0.569,0.000,0.431}70.5 & \color[rgb]{0.391,0.000,0.609}68.8 \\
Qwen3.6-plus & \color[rgb]{0.671,0.000,0.329}50.5 & \color[rgb]{0.864,0.000,0.136}62.7 & \color[rgb]{0.475,0.000,0.525}65.3 \\
Kimi-K2.6 & \color[rgb]{0.789,0.000,0.211}45.7 & \color[rgb]{0.881,0.000,0.119}62.3 & \color[rgb]{1.000,0.000,0.000}43.4 \\
GPT-5.1-Codex & \color[rgb]{1.000,0.000,0.000}37.1 & \color[rgb]{1.000,0.000,0.000}59.1 & \color[rgb]{0.396,0.000,0.604}68.6 \\
\bottomrule
\end{tabular}
}
    \vspace{-1em}
\end{wraptable}

To test whether the precision gap observed on \pdbhard{} generalizes to multi-line bugs, we evaluate the three generator models and the most recent models on \pdbmulti{}.
Table~\ref{tab:pdb_multi_results} shows that \pdbmulti{} is harder than \pdbhard{}, but \textbf{increasing bug granularity does not close the precision gap between models}.
The ranking is preserved: Claude-Opus-4.7 and Gemini-3.1-Pro again behave as relatively precise debuggers, with precision above $77\%$, while GPT-5.5 achieves a unit score above $84\%$ and recall comparable to the top models ($83\%$), but remains below $61\%$ precision.

\begin{figure*}[t]
    \centering
    \includegraphics[width=\linewidth]{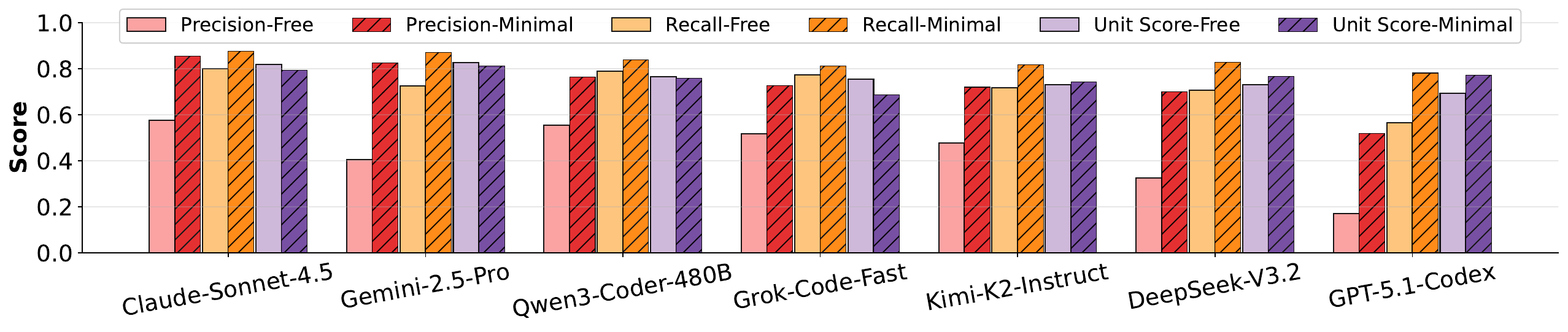}
    \caption{Comparison of model performance under minimal-debug and freeform prompting on a subset of \pdbfull. Freeform prompting leads to substantial drops in precision and recall across all models, indicating prompt-level constraints are necessary to increase debugging precision.}
    \label{fig:ablation_freeform}
\end{figure*}

\subsection{Prompting and data generation}
For ablation study, we randomly sample 500 examples from \pdbhard, to analyze how prompting and data generation strategies affect model debugging behavior.

\noindent\textbf{Freeform \emph{vs.} minimal debugging.}
In our main experiments, models are explicitly instructed to perform debugging with minimal edits. 
To assess the impact of this constraint, we conduct an ablation in which models are instead prompted to debug freely, without any restriction on edit scope (see Appendix~\ref{appsec:prompt} for prompts). 
Figure~\ref{fig:ablation_freeform} compares freeform to minimal-debug prompts.
Across all evaluated models, freeform prompting results in a substantial drop in edit-level precision and bug-level recall.
Even the strongest models, including Claude-Sonnet-4.5 and Qwen3-Coder-480B, achieve less than $60\%$ precision under freeform prompting.
Gemini-2.5-Pro exhibits a $40\%$ absolute drop in precision, indicating that its apparent debugging precision largely stems from instruction following rather than intrinsic edit awareness.
GPT-5.1-Codex performs particularly poorly under freeform prompts, failing to reach $20\%$ precision.
These results reinforce the regeneration behavior discussion, and demonstrate that \emph{prompt-level constraints are necessary but insufficient}: while minimal-debug prompts reduce over-editing, they do not fundamentally change underlying model behavior.

\noindent\textbf{Regeneration \emph{vs.} contamination.}
Although regeneration dominates model behavior on \pdb, an open question is whether this tendency is driven by data contamination, \ie, overlap between solutions and model pretraining data.
To disentangle these effects, we conduct two controlled analyses.
First, we rewrite ground-truth solutions using rewriter models (Claude-Sonnet-4.5 or GPT-5.1-Codex), producing semantically equivalent but surface-diverse references.
Second, we generate buggy programs using either the same model as the rewriter or a different generator model.
As shown in Table~\ref{tab:rewrite}, rewriting ground-truth solutions consistently makes debugging slightly easier, improving edit-level precision by $2.8$-$3.5\%$ on average.
In contrast, when buggy programs are generated by a different model from the rewriter, performance degrades, with unit-test pass rates dropping by up to $1.4\%$.
This indicates that cross-model generation introduces additional variability that is more difficult for models to resolve.
Taken together, these results suggest that while \emph{data contamination} may marginally influence debugging performance, it does not account for the pervasive regeneration behavior observed on \pdb.

\noindent\textbf{Bug generator analysis.} Table~\ref{tab:source_bug_gen} reveals that bugs generated by GPT-5.1-Codex are consistently the easiest to debug, while those generated by Claude-Sonnet-4.5 are the hardest. This ordering is consistent across all evaluated metrics.

\begin{table*}[t]
\centering
\begin{minipage}[t]{0.49\textwidth}
\centering
\tabcolsep 9pt
\caption{Rewriting ground truth consistently improves debugging precision, while using different generators makes buggy examples harder to solve.}
\label{tab:rewrite}
\resizebox{\linewidth}{!}{
\begin{tabular}{@{}lrrr@{}}
\toprule
\bf Data & \bf Precision  & \bf Recall & \bf Unit (\%) \\
\midrule
Raw 
& 73.0 
& 83.1 
& 76.2 \\
Rewrite-Same-Gen  
& \textbf{76.5} 
& 86.6 
& \textbf{76.4} \\
Rewrite-Different-Gen 
& 75.8 
& \textbf{86.8} 
& 74.8 \\
\bottomrule
\end{tabular}
}

\end{minipage}
\hfill
\begin{minipage}[t]{0.49\textwidth}
\centering
\tabcolsep 6pt
\caption{Precision, recall, and unit score comparison across source bug generation models for \pdbhard, averaged over debug models. }
\label{tab:source_bug_gen}
\resizebox{\linewidth}{!}{
\begin{tabular}{@{}lcrrr@{}}
\toprule
\textbf{BugGen Model} & \textbf{Count} & \textbf{Precision} & \textbf{Recall} & \textbf{Unit (\%)} \\
\midrule
GPT-5.1-Codex
& 1809
& 61.5
& 83.1
& 78.8 \\
Gemini-2.5-Pro
& 1937
& 58.1
& 75.7
& 71.0 \\
Claude-Sonnet-4.5
& 1988
& 49.9
& 67.6
& 67.8 \\
\bottomrule
\end{tabular}
}

\end{minipage}
\end{table*}

\subsection{Metric verification and error analysis}
\label{sec:metric_verf}
We conduct a qualitative error analysis by manually inspecting two categories of failures: (1) cases passing unit tests with imperfect precision or recall, and (2) cases failing unit tests despite containing partially correct edits. This analysis assesses the robustness of our precision and recall metrics. We randomly selected 240 examples from these categories and derived the taxonomy described below; detailed examples are provided in Appendix~\ref{appsec:debugcate}.

\noindent\textbf{Passing unit tests with imperfect precision or recall.}
In this category, models successfully resolve the intended bug but introduce extraneous modifications. 

\noindent\textit{Precision Analysis:} In 83.5\% of cases where unit tests pass, the recall score is also 1, but precision$<$$1$. We observe that 9.8\% of edits add redundant guard checks, 66.8\% modify correct code blocks, 13.7\% apply correct but non-minimal edits, and 7.8\% fully regenerate the solution. Notably, the remaining 1.9\% of patches have low precision because they fix bugs that were missing from the ground-truth solutions. 
Thus, our \textit{edit-level precision} accurately captures unnecessary edits.
Further, to assess human agreement on precision, annotators with over five years of Python experience rate each patch on a 1--5 scale, with higher scores indicating more targeted and intent-preserving repairs. \pdb{} precision strongly agrees with expert judgments, achieving Spearman $\rho = 0.72$.

\noindent\textit{Recall Analysis:} Conversely, 16.5\% of passing examples exhibit imperfect recall. In this scenario, we find that 70\% of examples are functionally correct but evade recall detection due to over-editing or structural rewrites. Furthermore, 10\% involve compounding bugs (\textit{only 1.65\% of all cases}), where an injected bug alters program logic such that other bugs change context, and another 20\% arise because a single bug allows for multiple minimal correct fixes. This indicates that, provided bug independence is maintained during dataset creation, our \textit{bug-level recall} is accurate for over 97.5\% of all the data.

\noindent\textbf{Failing unit tests with partially correct edits.}
In this category, models apply some correct edits but fail to fully resolve existing bugs or introduce new ones. We classify these failures into three types:
(1) \textit{Under-repair} (recall$<$$1$, precision$=$$1$): The model fixes some bugs without unnecessary edits but fails to apply all fixes (31.4\%).
(2) \textit{Imprecise repair} (recall$<$$1$, precision$<$$1$): The model both misses fixes and introduces unnecessary or harmful edits (29.4\%).
(3) \textit{Regressive repair} (recall$=$$1$): The model fixes all original bugs but introduces new errors that cause unit tests to fail; this accounts for the majority  (39.2\%).
This gap highlights a silent reasoning challenge unique to debugging: models must understand program structure and preserve intent while restoring functional correctness, rather than merely generating working code.

\section{Related Works}

Debugging is a critical yet time-consuming stage of the software development lifecycle~\citep{glass2002facts}, making it a natural target for automation with LLMs. 
Recent systems increasingly emulate real-world debugging workflows, achieving improved performance through hierarchical multi-agent architectures~\citep{lee2025unidebugger, bouzenia2025repairagent} and agent-based data synthesis with explicit communication~\citep{yang2025coast}. 
To evaluate such approaches, several general-purpose debugging benchmarks have been introduced, including those mined from historical bug-fixing commits~\citep{tian2024debugbench} and those expanding coverage across programming languages~\citep{liu2024mdeval}. 
Other benchmarks focus on specific dimensions of debugging, such as code editing~\citep{guo2024codeeditorbench}, systematic analyses of automated bug fixing~\citep{sobania2023analysis}, or broader coverage of debugging scenarios~\citep{yuan2025debug, huang-etal-2025-mldebugging, chai2024mceval}.

However, existing benchmarks predominantly rely on unit-test–based evaluation, which rewards models equally for rewriting large portions of code and for making minimal, targeted fixes. 
In contrast, \pdb introduces edit-level precision and bug-level recall, exposing fundamental shortcomings in current debugging systems and aligning evaluation more closely with real-world practices.
\section{Discussion}

We show that frontier LLMs often pass unit tests while remaining far from precise debugging, frequently relying on solution regeneration rather than targeted edits. By introducing \pdb{} with edit-level precision and bug-level recall, we expose behaviors that unit-test-only evaluation misses, revealing a substantial gap between functional correctness and genuine fault localization.
Results on \pdbhard{} and \pdbmulti{} show that improving debugging requires rethinking both evaluation and post-training objectives. More broadly, because both older and recent models struggle to precisely fix self-generated bugs under \pdb{}, the framework can serve not only as an evaluation benchmark but also as infrastructure for closing the training loop in self-improving coding models. Its precision-aware metrics provide reward signals focused on fault localization and edit minimality, while its plug-and-play pipeline enables scalable, controlled construction of debugging data from existing coding datasets.
\begin{ack}
We thank Yuqing Yang, Johnny Tian-Zheng Wei, Deqing Fu, Ting-Yun Chang, and Muru Zhang for reviewing early drafts of this work.
We also thank Haozhe Lu for valuable discussions and brainstorming.
WN was supported in part by the National Science Foundation under Grant No.\ CMMI-2427856.
RJ was supported in part by the National Science Foundation under Grant No. IIS-2403436. Any opinions, findings, and conclusions or recommendations expressed in this material are those of the author(s) and do not necessarily reflect the views of the National Science Foundation.
\end{ack}

\bibliographystyle{abbrvnat}
\bibliography{custom}

\clearpage
\appendix
\section*{Appendix}
\startcontents[appendix]
\renewcommand{\thesection}{\Alph{section}} 
\printcontents[appendix]{}{1}{\setcounter{tocdepth}{3}}
\setcounter{section}{0}

\clearpage

\section{Additional Related Works}
\label{appsec:related}

We discuss additional related work in the broader context of code generation and debugging.

\paragraph{Code Generation.} The capability of LLMs in code generation has been transforming both academia and industry. Beginning with seminal models like Codex~\citep{chen2021evaluating}, the field has rapidly advanced with the introduction of dozens of powerful code-centric models including Code Llama~\citep{roziere2023code}, StarCoder~\citep{li2023starcoder, allal2023santacoder}, CodeGen~\citep{nijkamp2022codegen}, CodeT5+~\citep{wang2023codet5+}, Qwen Coder Series \citep{hui2024qwen2,qwen3-coder}, and more recent GPT-5.1 and 5.2 codex \cite{gpt-5.1-codex, gpt-5.2-codex}. These models, trained on vast web-scale datasets of code, excel at synthesizing end-to-end programs from natural language prompts. To evaluate their capabilities, numerous benchmarks have been established, ranging from function-level synthesis tasks like HumanEval~\citep{chen2021evaluating} and CoNaLa~\citep{yin2018learning}, to more complex challenges including APPS~\citep{hendrycks2021measuring}, CodeContests~\citep{li2022competition}, SPOC~\citep{kulal2019spoc}, and BigCodeBench~\citep{zhuo2024bigcodebench}.
Beyond functional correctness, works like Vibe Checker~\citep{zhong2025vibe} and NoFunEval~\citep{singhal2024nofuneval} target the evaluation of models’ non-functional instruction-following abilities.
Recent agent-based systems like SWE-Agent~\citep{yang2024swe} and AutoGen~\citep{sevenhuijsen2025vecogen} demonstrate the potential of LLMs in autonomous development workflows.
Recently, \citet{zhang2025memorize} examined memorization effects in LLM-based code generation. We evaluated this hypothesis through targeted rewriting experiments (Table~\ref{tab:rewrite}) and found that memorization is not the root cause of regenerator-style behavior in debugging.

\paragraph{Debugging Frameworks.} As a critical and often time-consuming task, debugging has naturally emerged as another target for automation using LLMs. This need is further magnified by the fact that code generation models themselves are a significant source of buggy and potentially vulnerable code~\citep{ni2023lever, jin2023inferfix, mohsin2024can, tambon2025bugs, liu2026beyond, gao2025decoding}. Consequently, a spectrum of approaches have been proposed to leverage these models for program repair. Before LLMs, automated program repair was predominantly search-based, with optimization criteria that include the size of the patch~\citep{le2019automated}; recent work revisits patch minimality directly, \eg, AdaPR combines bug localization with preference learning to generate patches with minimal modifications~\citep{dai2025less}. Early work like Break-It-Fix-It~\citep{yasunaga2021break} ~\citep{zhang2023survey} introduced unsupervised learning for program repair, while CoText~\citep{phan2021cotext} explored multi-task learning with code-text transformers. Recent systems emulate real-world debugging workflows through sophisticated agent architectures: FixAgent~\citep{lee2025unidebugger} employs hierarchical multi-agent frameworks, RepairAgent~\citep{bouzenia2025repairagent} demonstrates autonomous repair capabilities, and COAST~\citep{yang2025coast} enhances debugging through communicative agent-based data synthesis. These approaches utilize techniques ranging from zero-shot prompting to multi-turn conversational agents~\citep{chen2021evaluating, wei2023copiloting, xia2024automated, zhong2024debug, islam2024llm, hu2026fitcache}.

\paragraph{Debugging Evaluation} To evaluate the performance of these LLM debugging approaches, a handful of benchmarks have been established. Early work like Defects4J~\citep{just2014defects4j} provided curated bug datasets from real-world Java projects, while recent benchmarks have adapted to the LLM era. \cite{tian2024debugbench} create debugging scenarios by mining historical bug-fixing commits. \cite{liu2024mdeval} curates multi-lingual code repair tasks spanning Python, Java, and JavaScript, while \cite{jain2024livecodebench} mitigates data contamination by using live programming contests. Specialized benchmarks like CodeEditorBench~\citep{guo2024codeeditorbench} focus on code editing capabilities, and analyses like \cite{sobania2023analysis} examine automatic bug fixing performance on existing datasets. A common limitation of these benchmarks, however, is their reliance on a simple, binary pass/fail metric on test cases~\citep{zhang2023critical}. Such coarse-grained evaluation is insufficient, as it cannot distinguish between a minimal, targeted fix and a complete code regeneration that merely passes the tests—a distinction crucial for understanding whether models truly comprehend debugging or simply regenerate working solutions. In contrast, our proposed \pdb disentangles debugging from code generation, introducing fine-grained evaluation metrics that assess not only functional correctness but also the precision, minimality, and human-like nature of code repairs, better reflecting real-world debugging practices where understanding and fixing the root cause is valued over wholesale replacement.

\paragraph{Mutation Testing and Fault Injection.}
Deliberately introducing faults into programs has a long history in software testing, under goals that differ from ours. \emph{Fault seeding} inserts known faults to evaluate a testing or quality-assurance process, and \emph{mutation testing}, its most widely studied form, applies small syntactic changes called \emph{mutation operators} to produce faulty program variants (\emph{mutants}) and measures test-suite adequacy by the fraction of mutants the tests detect~\citep{jia2011analysis}. Whether mutants are valid substitutes for real faults has been studied extensively~\citep{andrews2005mutation, ojdanic2023syntactic}, and practical tools such as PIT for Java~\citep{coles2016pit} implement standard operators, \eg, relational and arithmetic operator replacement. \emph{Fault injection}, in turn, introduces faults or fault conditions into a running system to study its fault-tolerance mechanisms~\citep{hsueh1997fault}. Bug injection in \pdb shares the mechanics of fault seeding but not its purpose: following prior debugging benchmarks~\citep{tian2024debugbench, liu2024mdeval}, we take the unit tests as the correctness oracle and use injected bugs as debugging tasks for a model to fix, rather than as probes of the tests themselves. This difference also sidesteps the \emph{equivalent-mutant problem}, in which a mutation leaves program behavior unchanged so that no test can detect it, which is undecidable in general~\citep{jia2011analysis}: \pdb retains an injected bug only if the unit tests fail on it, and the atomicity filter further requires every partial fix to fail (\S\ref{sec:pipeline}), so every retained bug is observably different from the ground-truth program.

We generate bugs with LLMs rather than mutation operators for three reasons. First, the generator is conditioned on the natural-language task description in addition to the code, so it produces faults that are plausible for the task being solved, such as a boundary mistake that a programmer implementing that specification might make, rather than uniformly sampled operator applications. Second, operator-style perturbations yield bugs that are too easy to separate models: in an early pilot, frontier models reached at least 90\% on both unit-test pass rate and precision on such perturbations, leaving little room for analysis or model comparison. Third, \pdbmulti requires contiguous multi-line bugs whose partial fixes still fail the tests, which are semantically coherent faults that operator-based seeding does not generate. Table~\ref{tab:mutation_odc} relates common mutation operators to the defect taxonomy below.

\paragraph{Orthogonal Defect Classification.}
The Orthogonal Defect Classification (ODC) framework~\citep{chillarege1992orthogonal} partitions software defects into eight semantically disjoint defect types---\textit{Function}, \textit{Interface}, \textit{Checking}, \textit{Assignment}, \textit{Timing/Serialization}, \textit{Build/Package/Merge}, \textit{Documentation}, and \textit{Algorithm}---so that any given bug can be assigned to exactly one type without overlap. \textit{Function} errors affect significant capability and require a formal design change, and \textit{Documentation} errors concern publications rather than code, so neither can be synthesized as a line-level edit inside a static seed program; we fold errors in interacting with external libraries into \textit{Build/Package/Merge} (Table~\ref{tab:odc_taxonomy}) instead of a separate \textit{Interface} type. We retain the five types that admit purely code-level expression---\textbf{Assignment}, \textbf{Checking}, \textbf{Algorithm}, \textbf{Build/Package/Merge}, and \textbf{Timing/Serialization}---and use them as the in-context taxonomy that conditions the bug-injection generator. We use ODC as a taxonomy rather than as an injection mechanism: its categories are language-agnostic, so adapting \pdb to another language only requires new per-category in-context examples, and they enable the category-level analysis in \S\ref{sec:results} (Figure~\ref{fig:bug_type_breakdown}).

\begin{table*}[t]
\centering
\small
\caption{ODC-style taxonomy of common programming defects with summarized descriptions. These are used as in-context examples.}
\label{tab:odc_taxonomy}
\resizebox{\textwidth}{!}{
\begin{tabular}{lll}
\toprule
\textbf{ODC Category} & \textbf{Sub-category} & \textbf{Brief Description} \\
\midrule

\multirow{6}{*}{\textbf{Assignment}}
& Mutability Trap
& Mutable default arguments cause unintended shared state across calls. \\
& Late Binding in Closures
& Loop variables captured by reference, yielding unexpected final values. \\
& List Multiplication Surprise
& List multiplication creates multiple references to the same inner object. \\
& Built-in Shadowing
& Assigning to names like \texttt{list} or \texttt{sum} hides built-ins. \\
& Variable Shadowing
& Inner-scope variables obscure outer-scope references. \\
& Name Error
& Variable is used before being assigned or defined. \\
\midrule

\multirow{8}{*}{\textbf{Checking}}
& Off-by-One Error
& Boundary condition is shifted by exactly one element or unit. \\
& Negation Error
& Boolean condition is logically inverted. \\
& Missing or Incomplete Checks
& Absent validation leads to runtime errors (e.g., KeyError, TypeError). \\
& Overwriting Built-in Names
& Built-in identifiers are reassigned, breaking later function calls. \\
& Variable Shadowing
& Confusing variable scope leads to incorrect condition evaluation. \\
& Chained Boolean Comparison Logic
& Misparsed chained comparisons yield unintended logic. \\
& Implicit Boolean Conversion
& Empty collections and \texttt{None} are conflated in boolean context. \\
& Membership Logic Flaws
& Misunderstanding how membership tests behave for data types. \\
\midrule

\multirow{6}{*}{\textbf{Algorithm}}
& Wrong Math Expression
& Mathematical formula or operands are incorrectly specified. \\
& Modifying While Iterating
& Collection is altered during iteration, skipping or misprocessing elements. \\
& Function Algorithm Misunderstanding
& Function behavior is misunderstood (e.g., substring vs. set semantics). \\
& Function Argument Misunderstanding
& Incorrect interpretation of function arguments or defaults. \\
& Infinite Loop / Recursion
& Termination condition is missing or unreachable. \\
& Other Logical Errors
& Deeper algorithmic invariants are violated during execution. \\
\midrule

\multirow{2}{*}{\textbf{Build/Package/Merge}}
& Invalid API Call
& Method is invoked on an unsupported data type or abstraction. \\
& Dependency Version Conflicts
& Code relies on APIs removed or changed across library versions. \\
\midrule

\multirow{2}{*}{\textbf{Timing/Serialization}}
& Serialization Issue
& Non-serializable objects are passed to pickle or JSON encoders. \\
& Async Blocking
& Blocking calls inside async code stall the event loop. \\
\bottomrule

\end{tabular}
}
\end{table*}

Table~\ref{tab:odc_taxonomy} expands each retained ODC category into the concrete sub-patterns we surface to the generator. \textit{Assignment} captures Python-specific reference and scoping pitfalls (mutable defaults, late binding in closures, shadowed built-ins); \textit{Checking} covers boundary, negation, and validation faults at branch points; \textit{Algorithm} groups deeper logical and arithmetic errors that violate program invariants; \textit{Build/Package/Merge} reflects API-misuse and dependency-version faults that surface when a program meets its environment; and \textit{Timing/Serialization} encodes serialization and async-blocking issues that are easy to introduce but hard to diagnose by inspection. Anchoring the generator to this taxonomy avoids degenerate failure modes where the model converges on a single defect family (typically \textit{Checking}) and instead yields broad coverage across the precision/recall regimes our evaluation later stratifies.

\paragraph{ODC Categories \emph{vs.} Mutation Operators.}
\begin{table*}[t]
\centering
\small
\caption{Default mutation operators of PIT~\citep{coles2016pit} and mutmut mapped to ODC categories and to the closest \pdb sub-category in Table~\ref{tab:odc_taxonomy}. Examples are written in Python. Operators concentrate on \textit{Checking} and on arithmetic within \textit{Algorithm}; the sub-categories in the last row have no default operator counterpart.}
\label{tab:mutation_odc}
\resizebox{\textwidth}{!}{
\begin{tabular}{lllll}
\toprule
\textbf{Mutation Operator} & \textbf{Tool} & \textbf{Example} & \textbf{ODC Category} & \textbf{Closest \pdb Sub-category} \\
\midrule

Relational boundary replacement & PIT, mutmut & \texttt{i < n} $\rightarrow$ \texttt{i <= n} & \textbf{Checking} & Off-by-One Error \\
Numeric constant change & mutmut & \texttt{nums[0]} $\rightarrow$ \texttt{nums[1]} & \textbf{Checking} & Off-by-One Error \\
Conditional negation & PIT, mutmut & \texttt{x == y} $\rightarrow$ \texttt{x != y}; \texttt{True} $\rightarrow$ \texttt{False} & \textbf{Checking} & Negation Error \\
Membership and identity swap & mutmut & \texttt{x in seen} $\rightarrow$ \texttt{x not in seen}; \texttt{is} $\rightarrow$ \texttt{is not} & \textbf{Checking} & Membership Logic Flaws; Negation Error \\
Logical connector replacement & mutmut & \texttt{a and b} $\rightarrow$ \texttt{a or b} & \textbf{Checking} & Chained Boolean Comparison Logic \\
\midrule

Arithmetic and bitwise replacement & PIT, mutmut & \texttt{a + b} $\rightarrow$ \texttt{a - b}; \texttt{a \& b} $\rightarrow$ \texttt{a | b} & \textbf{Algorithm} & Wrong Math Expression \\
Increment and sign change & PIT, mutmut & \texttt{i += 1} $\rightarrow$ \texttt{i -= 1}; \texttt{-x} $\rightarrow$ \texttt{+x} & \textbf{Algorithm} & Wrong Math Expression; Infinite Loop / Recursion \\
Argument replacement or removal & mutmut & \texttt{f(x, k)} $\rightarrow$ \texttt{f(None, k)} & \textbf{Algorithm} & Function Argument Misunderstanding \\
String-method swap & mutmut & \texttt{s.find(c)} $\rightarrow$ \texttt{s.rfind(c)} & \textbf{Algorithm} & Function Algorithm Misunderstanding \\
Control-flow change or call removal & PIT, mutmut & \texttt{continue} $\rightarrow$ \texttt{break}; \texttt{nums.sort()} removed & \textbf{Algorithm} & Other Logical Errors \\
\midrule

Copy-depth swap & mutmut & \texttt{deepcopy(grid)} $\rightarrow$ \texttt{copy(grid)} & \textbf{Assignment} & List Multiplication Surprise (shared inner objects) \\
Value replacement & PIT, mutmut & \texttt{return total} $\rightarrow$ \texttt{return None} & \textbf{Assignment} & --- \\
\midrule

\multirow{3}{*}{No default operator} & \multirow{3}{*}{---} & \multirow{3}{*}{---} & \textbf{Assignment}, \textbf{Checking}, & Mutability Trap, Late Binding in Closures, Built-in Shadowing, Variable Shadowing, Name Error, \\
& & & \textbf{Algorithm}, \textbf{Build/Package/Merge}, & Missing or Incomplete Checks, Overwriting Built-in Names, Implicit Boolean Conversion, \\
& & & \textbf{Timing/Serialization} & Modifying While Iterating, Invalid API Call, Dependency Version Conflicts, Serialization Issue, Async Blocking \\
\bottomrule

\end{tabular}
}
\end{table*}

Table~\ref{tab:mutation_odc} maps the default mutation operators of PIT and mutmut to ODC categories and to the closest \pdb sub-category in Table~\ref{tab:odc_taxonomy}. The operators concentrate on \textit{Checking}, through boundary, negation, and membership flips, and on arithmetic within \textit{Algorithm}; together they reach only 10 of the 23 distinct sub-categories. The remaining 13, including Python-specific reference and scoping pitfalls under \textit{Assignment} and environment-level faults under \textit{Build/Package/Merge} and \textit{Timing/Serialization}, have no default operator counterpart. Conditioning an LLM generator on the ODC taxonomy therefore covers defect types that operator-based seeding does not reach, while still subsuming the operator-style faults within \textit{Checking} and \textit{Algorithm}.

\clearpage

\section{Additional Experiments}
\label{appsec:exps}

\S\ref{appsec:pdbfull} shows that the model ranking observed on \pdbhard{} persists on the unfiltered \pdbfull{} pool. \S\ref{appsec:length}--\S\ref{appsec:rewrite} report ablations that isolate the contribution of buggy code length and source rewriting to the headline precision/recall numbers; the defect-category breakdown appears in \S\ref{sec:results}. \S\ref{appsec:debugbench}--\S\ref{appsec:contamination} test transfer to a human-curated benchmark and a contemporaneous bug generator, addressing the concern that our findings are confined to a particular synthetic distribution.

\subsection{Results on the full \pdbfull{} pool}
\label{appsec:pdbfull}

\begin{table}[ht]
\centering
\tabcolsep 5pt
\caption{Precision, recall, and unit score on the \pdbfull set. Blue indicates better performance, while red indicates worse.}
\label{tab:model_performance_pdb_full}
\begin{tabular}{@{}lrrr@{}}
\toprule
\bf Model & \bf Precision  & \bf Recall & \bf Unit (\%) \\
\midrule
Claude-Sonnet-4.5 & \color[rgb]{0.000,0.000,1.000}\textbf{78.1$\pm$0.7} & \color[rgb]{0.127,0.000,0.873}85.7$\pm$0.6 & \color[rgb]{0.176,0.000,0.824}81.9$\pm$0.9 \\
Gemini-2.5-Pro & \color[rgb]{0.008,0.000,0.992}77.9$\pm$0.7 & \color[rgb]{0.000,0.000,1.000}\textbf{87.5$\pm$0.6} & \color[rgb]{0.067,0.000,0.933}83.8$\pm$0.8 \\
Qwen3-Coder-480B & \color[rgb]{0.156,0.000,0.844}73.5$\pm$0.8 & \color[rgb]{0.341,0.000,0.659}82.4$\pm$0.7 & \color[rgb]{0.424,0.000,0.576}77.4$\pm$0.9 \\
Kimi-K2-Instruct & \color[rgb]{0.427,0.000,0.573}65.8$\pm$0.8 & \color[rgb]{0.570,0.000,0.430}78.8$\pm$0.7 & \color[rgb]{0.667,0.000,0.333}73.0$\pm$1.0 \\
Grok-Code-Fast & \color[rgb]{0.512,0.000,0.488}63.8$\pm$0.9 & \color[rgb]{1.000,0.000,0.000}73.2$\pm$0.8 & \color[rgb]{1.000,0.000,0.000}67.1$\pm$1.1 \\
Kimi-K2-Thinking & \color[rgb]{0.605,0.000,0.395}61.3$\pm$0.8 & \color[rgb]{0.423,0.000,0.577}81.2$\pm$0.7 & \color[rgb]{0.226,0.000,0.774}80.8$\pm$0.9 \\
DeepSeek-V3.2 & \color[rgb]{0.702,0.000,0.298}58.6$\pm$0.9 & \color[rgb]{0.788,0.000,0.212}76.2$\pm$0.8 & \color[rgb]{0.368,0.000,0.632}78.2$\pm$0.9 \\
DeepSeek-V3.2-Thinking & \color[rgb]{0.803,0.000,0.197}56.0$\pm$0.9 & \color[rgb]{0.733,0.000,0.267}77.5$\pm$0.8 & \color[rgb]{0.000,0.000,1.000}\textbf{84.7$\pm$0.8} \\
GPT-5.1-Codex & \color[rgb]{1.000,0.000,0.000}50.3$\pm$0.8 & \color[rgb]{0.727,0.000,0.273}77.8$\pm$0.8 & \color[rgb]{0.156,0.000,0.844}82.0$\pm$0.9 \\
\bottomrule
\end{tabular}
\end{table}

Table~\ref{tab:model_performance_pdb_full} reports the same nine models on \pdbfull{}, the unfiltered superset from which \pdbhard{} is sampled. All three metrics are 4--8\% higher than on \pdbhard{} (Table~\ref{tab:model_performance}), as expected: the difficulty filter that defines \pdbhard{} removes easy examples that would otherwise lift averages. Crucially, the model ordering is preserved: Claude-Sonnet-4.5 and Gemini-2.5-Pro retain the highest precision (78.1\% and 77.9\%), Qwen3-Coder-480B remains the strongest mid-tier debugger, and GPT-5.1-Codex remains the weakest on precision (50.3\%) despite a competitive unit-test pass rate. The stability of this ranking across the full and filtered pools indicates that the precision gap reported in the main paper is a property of model behaviour, not an artefact of the difficulty subset.

\begin{figure*}[ht]
    \centering
    \begin{subfigure}[t]{0.49\linewidth}
        \centering
        \includegraphics[width=\linewidth]{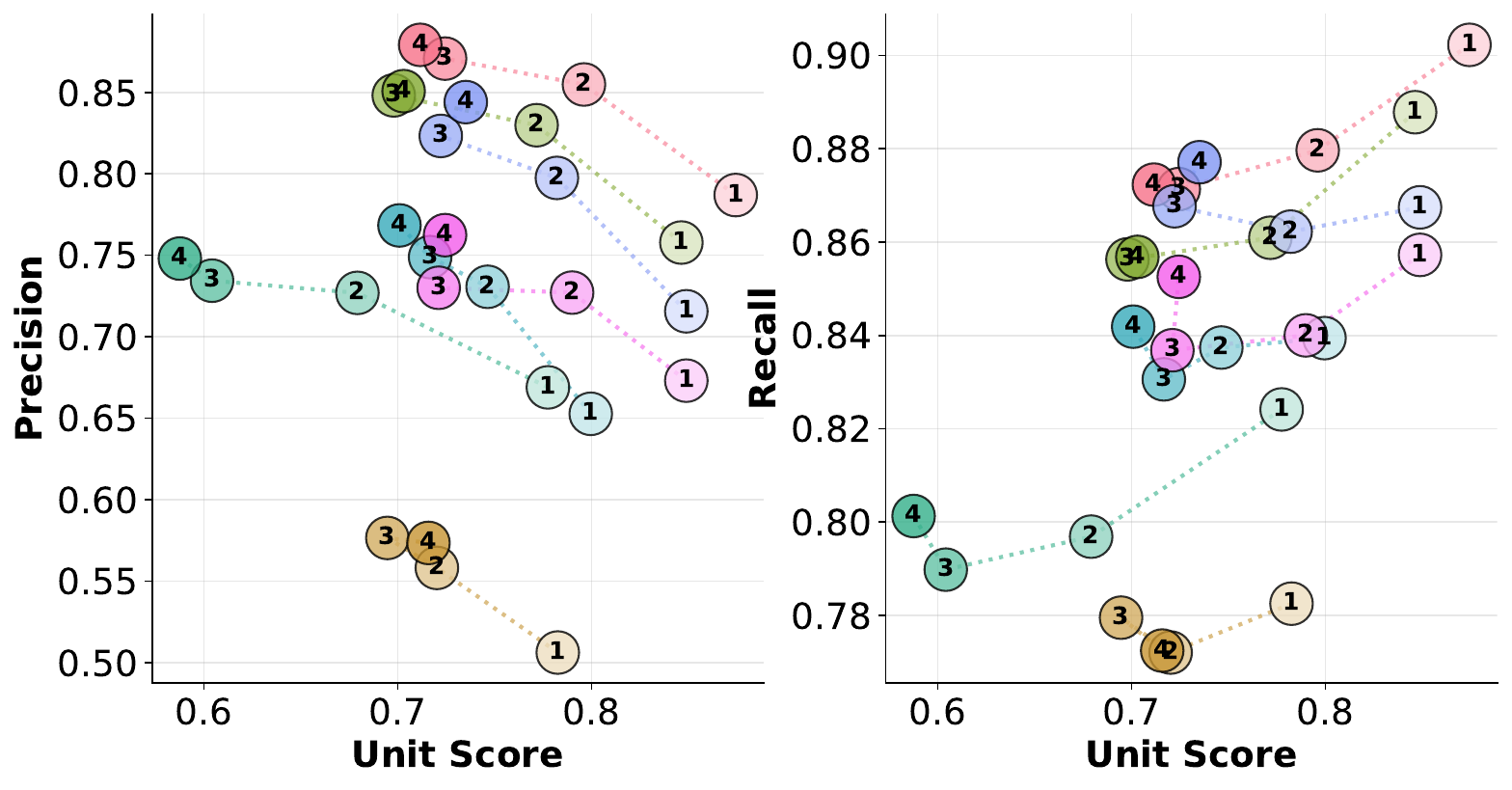}
    \end{subfigure}
    \hfill
    \begin{subfigure}[t]{0.49\linewidth}
        \centering
        \includegraphics[width=\linewidth]{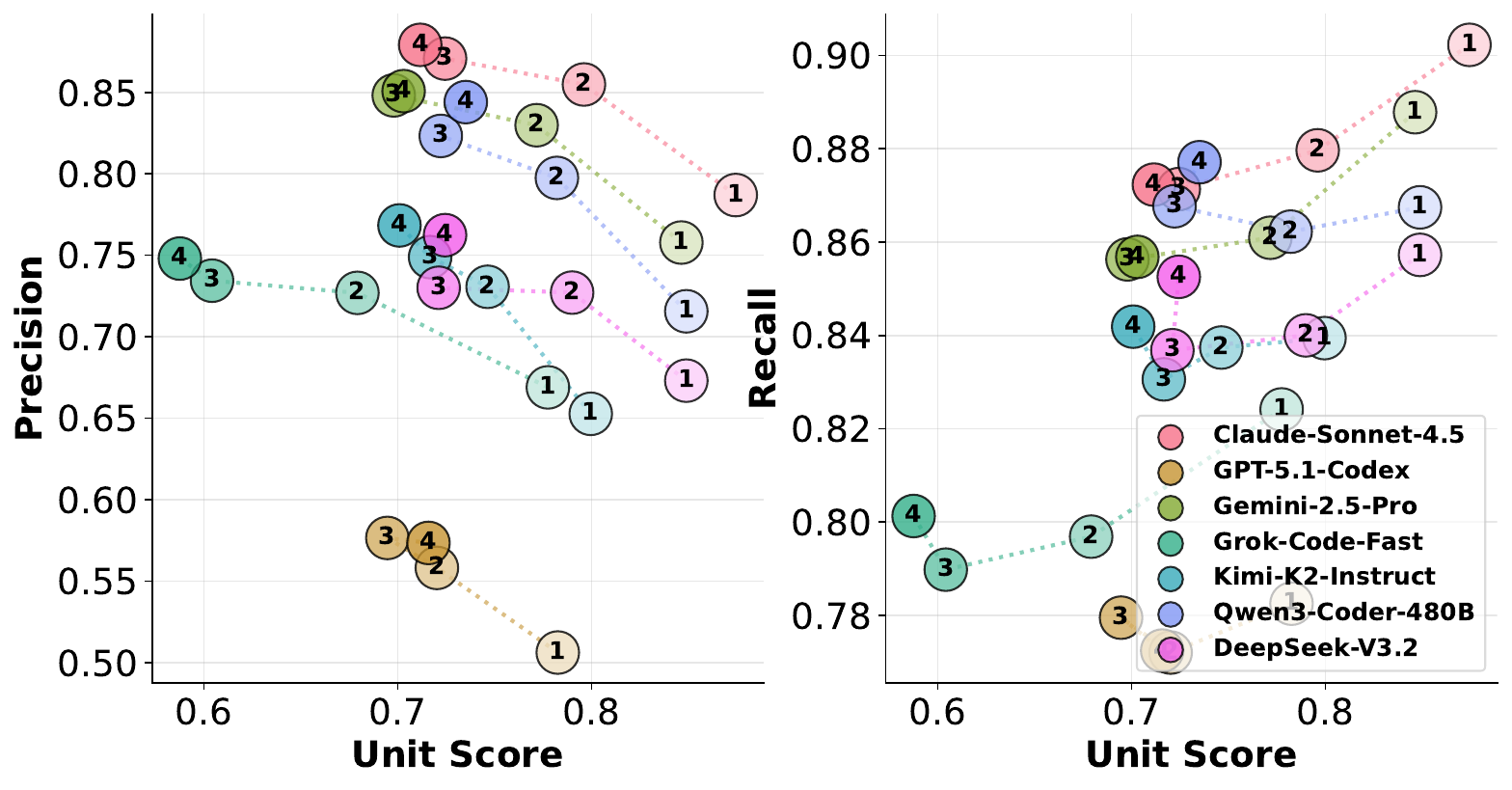}
    \end{subfigure}
    \caption{Per-bug-count breakdown of precision, recall, and unit score on the full \pdbfull{} pool drawn from \bigcode{} (left) and \livecode{} (right). Bug counts $k\in\{1,2,3,4\}$ are annotated next to each marker. Larger $k$ pulls precision down monotonically while unit-test pass rate decays more slowly; recall is essentially flat on \bigcode{} and increases with $k$ on \livecode{}.}
    \label{fig:bug-count-breakdown-all}
\end{figure*}

Figure~\ref{fig:bug-count-breakdown-all} stratifies the same metrics by the number of injected bugs $k$ on each source benchmark. The negative correlation between unit score and precision sharpens compared to the filtered \pdbhard{} version (Figure~\ref{fig:bug-count-breakdown}): every additional bug increases the chance that the model touches at least one needed line, but it also magnifies its over-editing tendency. The recall trend separates the two source benchmarks. On the API-heavy \bigcode{}, per-bug fix difficulty is roughly stationary, so recall stays within a 5\% band across $k$. On the algorithm-heavy \livecode{}, individual bugs become harder when the program is already broken in multiple places, and recall tracks the unit-test pass rate upward as $k$ grows. Both observations were already visible on \pdbhard{} and become more pronounced on the larger \pdbfull{} pool, indicating they are systematic rather than sample-specific.

\subsection{Buggy code length scaling}
\label{appsec:length}

\begin{figure}[h]
    \centering
    \includegraphics[width=0.7\linewidth]{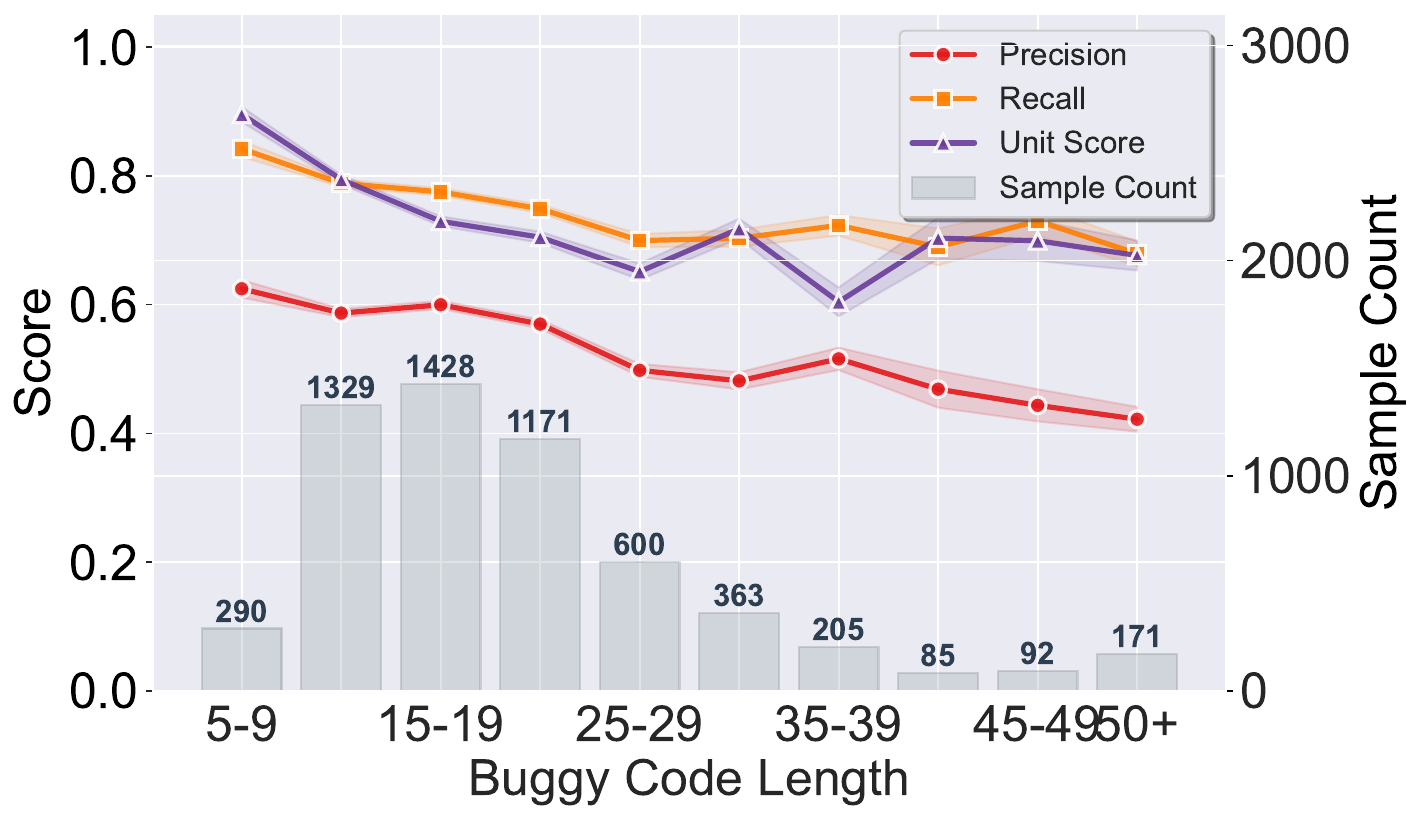}
    \caption{Model-averaged precision, recall, and unit-test pass rate on \pdbhard{}, plotted as a function of buggy code length (lines of code). All three metrics decay together as the program grows.}
    \label{fig:ablation_length}
\end{figure}

Figure~\ref{fig:ablation_length} averages each metric across the nine evaluated models and bins the results by the length of the buggy program. 
Longer programs hurt every metric simultaneously: the model first has to localize the bug among more candidate lines, then has to apply a precise edit without disturbing the surrounding code. 
Because the curves move in lockstep, the precision drop is not explained by tasks becoming purely harder to solve; if that were the case, recall and unit score would degrade while precision held steady. 
Instead, longer programs invite proportionally more spurious edits, which is consistent with the regeneration-style behaviour identified in the main text.

\subsection{Source rewriting and cross-model bug generation}
\label{appsec:rewrite}

\begin{figure*}[h]
    \centering
    \includegraphics[width=\linewidth]{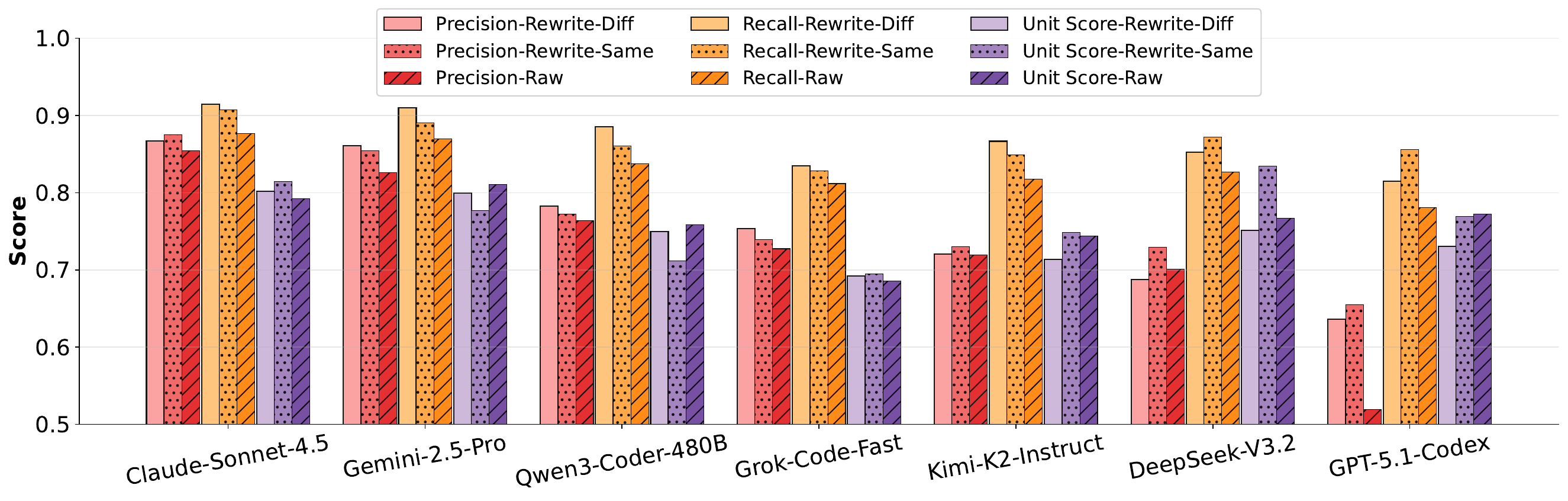}
    \caption{Per-model precision, recall, and unit-test pass rate on \pdbhard{} when ground-truth solutions are rewritten by the same generator that introduced the bugs (\textit{Same-Gen}) versus a different generator (\textit{Different-Gen}), compared against the unmodified raw data.}
    \label{fig:ablation_rewrite}
\end{figure*}

Figure~\ref{fig:ablation_rewrite} expands the aggregate rewrite ablation in Table~\ref{tab:rewrite} into a per-model view. The aggregate finding---rewriting raises precision by 2.8--3.5\% on average, while cross-model bug generation lowers unit pass rate by up to 1.4\%---holds for every model individually, but the magnitude varies. Models with the strongest baseline precision (Claude-Sonnet-4.5, Gemini-2.5-Pro) gain the least from rewriting, indicating that they were not relying on incidental string overlap with their own pretraining data; conversely, the weakest models on precision (GPT-5.1-Codex, DeepSeek-V3.2) gain disproportionately less than they lose under cross-model generation, the opposite of what a memorization-driven account would predict. The breakdown therefore reinforces the conclusion in \S\ref{sec:results} that contamination is at most a secondary contributor to the observed precision gap.

\subsection{Real-world debugging on \textsc{DebugBench}}
\label{appsec:debugbench}

\begin{table}[t]
\centering
\tabcolsep 12pt
\caption{Model performance on DebugBench, evaluated with the \pdb framework. The qualitative pattern observed on \pdbsh and \pdbmulti persists: high unit-test pass rates coexist with substantially lower edit-level precision.}
\label{tab:debugbench_results}
\begin{tabular}{@{}lrrr@{}}
\toprule
\bf Model & \bf Precision & \bf Recall & \bf Unit (\%) \\
\midrule
Claude-Sonnet-4.5   & 78.4          & 87.3          & 87.5 \\
Gemini-2.5-Pro      & \textbf{79.4} & \textbf{89.4} & 85.0 \\
GPT-5.1-Codex       & 61.9          & 74.0          & \textbf{90.0} \\
\bottomrule
\end{tabular}
\end{table}

Because PDB constructs debugging instances by injecting verified bugs, a natural concern is whether the observed precision gap is an artifact of synthetic bug generation. We examine this concern from two complementary perspectives: whether the same model behavior transfers to naturally occurring bugs, and whether PDB instances are distinguishable from real bugs at the instance level.

First, we extend the \pdb{} evaluation framework to \textsc{DebugBench}~\citep{tian2024debugbench}, a human-validated synthetic debugging benchmark, to test whether the precision/recall pattern transfers. \textsc{DebugBench} does not annotate bug counts, so we approximate them through edit structure: we keep only examples whose ground-truth fixes form contiguous edit blocks under the same stride constraint $s=5$ used for \pdbmulti{}, treating each block as one bug. The filter yields a subset of 40 examples, each containing exactly one bug under the stride-5 grouping, sampled uniformly at random from the qualifying pool.

Table~\ref{tab:debugbench_results} reports three frontier models evaluated under our protocol on this subset. Absolute scores are higher than on \pdbhard{} or \pdbmulti{}, reflecting that \textsc{DebugBench} bugs are individually easier (one block per example, drawn from real fixes that human reviewers already accepted). Even so, the qualitative pattern persists: GPT-5.1-Codex obtains the highest unit-test pass rate (90.0\%) but the lowest precision (61.9\%), with a recall--precision gap of roughly 12 points; Claude-Sonnet-4.5 and Gemini-2.5-Pro again sit in the precise-debugger regime with precision approaching 80\%. The same recall--precision dissociation that motivates \pdb{} therefore reappears on a benchmark that we did not construct, indicating that the gap is a property of how current models repair code rather than of any specific synthetic injection pipeline.

We further conduct a controlled human study. We sample 100 PDB instances from the LiveCodeBench subset ~\citep{jain2024livecodebench}, evenly stratified across bug counts $k \in \{1,2,3,4\}$, and 100 \textsc{DebugBench} instances matched by bug count and problem domain. Expert judges with more than five years of Python experience are shown only the task description and buggy program, with the instances randomly shuffled and their provenance concealed. For each instance, judges predict whether the bug occurred naturally or was injected by an LLM. Their classification accuracy is 55.0\% (Wilson 95\% CI: 48.1\%--61.7\%), which is not significantly different from the 50\% chance level under a two-sided exact binomial test ($p=0.179$). We therefore find no evidence that experts can reliably distinguish PDB bugs from naturally occurring bugs. 

\subsection{Unit-test oracle reliability}
\label{appsec:unit-test}
First, the source benchmarks were themselves built with test adequacy as an explicit design goal. BigCodeBench\cite{zhuo2024bigcodebench} reports an average branch coverage of 99\% per task, with multiple test cases exercising API-call behavior. LiveCodeBench \cite{jain2024livecodebench} draws from live competitive-programming contests, whose tests are adversarially designed to eliminate incorrect solutions, and further augments them with generator-based adversarial tests. SWE-smith\cite{yang2025swe} inherits the full test suites of real repositories and validates each instance via \texttt{FAIL\_TO\_PASS} tests (the bug must break them) and \texttt{PASS\_TO\_PASS} tests (the rest of the suite must still pass), so both bug detection and non-regression are checked against production test suites.

Second, the remaining risk is false positives at scoring time: a model patch that passes the tests while being actually incorrect. To quantify this, we randomly sampled 100 examples in which the model's patch passed the unit tests, and asked expert annotators with more than five years of Python experience to independently verify each patched program against the task description. 0 of 100 patches were found to be erroneous, indicating that test-passing is a reliable correctness signal on our benchmarks.

Third, we measure the false-positive rates from the patches on PDB-Single (LiveCodeBench) that pass the inherited tests but differ from the ground truth, the entire population in which a scoring false positive could occur. We sampled two disjoint batches of 108 patches (24 per model in total for the models reported in \ref{tab:model_performance}) and submitted all 216 to LeetCode's official online judge, which runs the full hidden test suites. Results: 212/216 accepted; 4 rejected, a false-positive rate of 1.85\%. The rejections were spread across four different models, with one rejection each for Gemini-2.5-Pro, Qwen3-Coder-480B, Kimi-K2-Thinking, and DeepSeek-V3.2. Note that in code completion tasks, even with human-LLM collaborative test generation\cite{ma2026rethinking}, LiveCodeBench has a false-positive rate of 6.5\% in a sample of 101 examples.

We hypothesize the rate is this low because debugging differs structurally from generation: candidate patches start from a nearly correct buggy program, so a weak suite only needs to catch the model's divergence from the reference, which is typically small unless the model substantially regenerates the program. Consistent with this, rejected patches have markedly lower precision than accepted ones (mean 0.25 vs. 0.45; 3 of 4 rejections below 0.5): heavily rewritten patches are exactly where hidden-test failures hide. We find this a striking additional argument for the precision metric itself: precision flags the patches that functional scoring may be unreliable.

\subsection{Multi-line bugs and post-release models on \pdbmulti{}}
\label{appsec:contamination}

A natural concern with \pdbhard{} is that the headline numbers in Table~\ref{tab:model_performance} characterize models that may have observed the seed BCB/LCB programs during pretraining or post-training. Five of the strongest contemporary models, Claude-Opus-4.7~\cite{claude-opus-4.7}, Gemini-3.1-Pro~\cite{gemini-3.1-pro}, GPT-5.5~\cite{gpt-5.5}, Qwen-3.6-Plus~\cite{qwen36plus}, and Kimi-K2.6~\cite{kimi-k2.6}, were released after \pdbhard{} became publicly available, so we cannot rule out that any improvement they show on the same data partly reflects contamination of the seed programs themselves.

To mitigate this concern, we use Claude-Opus-4.7, a contemporaneous frontier model that did not exist when the original generators ran, to synthesize multi-line, repository-scale bugs and assemble \pdbmulti{}. Table~\ref{tab:pdb_multi_results} reports the five post-release models alongside three of the original generators on \pdbmulti{}. The same precision deficit reappears: Claude-Opus-4.7 and Gemini-3.1-Pro lead with precision near 78\%, GPT-5.5 attains a competitive unit pass rate (84.3\%) but only 60.9\% precision, and GPT-5.1-Codex collapses to 37.1\% precision despite a 68.6\% unit-test pass rate. Two implications follow. First, the precision gap is robust to the choice of bug generator and to whether the seed code was previously seen by the debugger. Second, because the \pdb{} pipeline can be re-instantiated with any newer frontier generator, the benchmark naturally evolves with the model frontier rather than aging into the training data.

\clearpage

\section{Algorithm on Precision and Recall}
\label{appsec:algo}

We formalize the block-matching primitives used in the precision and recall metrics defined in \S\ref{sec:setup}. Algorithm~\ref{alg:map} defines the matcher $\map$, which pairs predicted edit blocks with ground-truth edit blocks; Algorithm~\ref{alg:map_epsilon} defines $\ess_\gU$, which verifies each pairing with unit tests and computes the $\epsilon$-relaxed essential edit size $(|\hat{E}_i|)_\epsilon$ used in Eq.~(\ref{eqn:precision_relax}). Both operate on edit blocks of any length, so they apply unchanged to the single-line bugs of \pdbhard and the multi-line bugs of \pdbmulti.

The matcher $\map$ aligns predicted edit blocks to ground-truth edit blocks in two passes. Pass~1 accepts exact matches without testing: a predicted block that is identical to a ground-truth block, covering the same lines with the same edits. Matching whole blocks ensures that a partial fix inside a multi-line ground-truth block is never mistaken for an exact match. Pass~2 matches each remaining predicted block by three progressively looser criteria: the block wraps the start of a ground-truth block (2.1); it occupies the same local context, measured by overlapping lines before and after both blocks (2.2); or it carries the same edits in the same order at a different location (2.3). Each Pass~2 pairing is recorded with a tester program that applies the predicted block in place of the matched ground-truth blocks, together with the ground-truth fixes for all other bugs. Because one predicted block may cover several ground-truth blocks, the output is a partial many-to-one matching; unmatched predicted blocks count against precision and unmatched ground-truth blocks count against recall.

The procedure $\ess_\gU$ first runs each tester, marking a pairing as semantically correct for its bugs if the tester passes $F_\gU$ (exact matches need no test). Each correct pairing on ground-truth blocks $\gG$ may then be credited with at most $\kappa(\gG)=\sum_{B^{\text{gt}}\in\gG}(|B^{\text{gt}}.\Delta|+\epsilon)$ lines, i.e., $|E_i|+\epsilon$ per matched bug $i$, which reduces to $\epsilon+1$ for single-line bugs. Finally, a deep redundancy check enumerates contiguous sub-blocks of the predicted block up to this cap, applies each to the buggy program with the ground-truth fixes for the other bugs, and keeps the smallest one that still passes the tests as the essential edit size. Extra predicted lines are thus credited only when the model's fix actually needs them, while wholesale regeneration receives no credit beyond the cap; the qualitative consequences are examined in \S\ref{sec:metric_verf}.

\begin{algorithm*}
\caption{$\operatorname{map}$: mapping predicted edits to ground-truth edits}
\label{alg:map}
\DontPrintSemicolon
\KwIn{Buggy program $C_b$; Predicted edits $\Delta_{\text{pred}}$; GT edits $\Delta_{\text{gt}}$}
\KwOut{Matched blocks $\mathcal{M}$}

\BlankLine
$\mathcal{M}\leftarrow\emptyset$, $\Delta^{\text{rem}}_{\text{gt}}\leftarrow \Delta_{\text{gt}}$, $\Delta^{\text{rem}}_{\text{pred}}\leftarrow \Delta_{\text{pred}}$ \tcp*{each element in $\Delta$ has fields: line, edit}
$\mathcal{B}^{\text{all}}_{\text{gt}}\leftarrow \textsc{ParseToBlocks}(\Delta^{\text{rem}}_{\text{gt}})$\;
$\mathcal{B}^{\text{rem}}_{\text{gt}}\leftarrow \textsc{ParseToBlocks}(\Delta^{\text{rem}}_{\text{gt}})$ \tcp*{each block in $\mathcal{B}$ has fields: start, end and $\Delta$}

\BlankLine
\textbf{Pass 1: Exact block-level matches (EM).}\;
\tcp{Whole blocks only: a partial fix of a multi-line GT block is not an EM.}
\ForEach{predicted block $B^{\text{pred}}\in \textsc{ParseToBlocks}(\Delta^{\text{rem}}_{\text{pred}})$ \textbf{in descending} start}{
    \If{$\exists B^{\text{gt}}\in \mathcal{B}^{\text{rem}}_{\text{gt}}$ \textbf{with} $(B^{\text{gt}}.\text{start}, B^{\text{gt}}.\text{end})=(B^{\text{pred}}.\text{start}, B^{\text{pred}}.\text{end})$ \textbf{and} $\textsc{Equal}(B^{\text{pred}}.\Delta, B^{\text{gt}}.\Delta)$}{
        $\mathcal{M} \leftarrow \mathcal{M} \cup \{\textsc{MakeMatch}(B^{\text{pred}}, \{B^{\text{gt}}\}, \text{none})\}$ \tcp*{No need to test}
        remove the lines of $B^{\text{pred}}$ from $\Delta^{\text{rem}}_{\text{gt}}$ and from $\Delta^{\text{rem}}_{\text{pred}}$\;
        remove $B^{\text{gt}}$ from $\mathcal{B}^{\text{rem}}_{\text{gt}}$\;
    }
}
$\mathcal{B}^{\text{rem}}_{\text{pred}}\leftarrow \textsc{ParseToBlocks}(\Delta^{\text{rem}}_{\text{pred}})$\;

\BlankLine
\textbf{Pass 2: Block-level matching.}\;
\For{$j\leftarrow 1$ \KwTo $|\mathcal{B}^{\text{rem}}_{\text{pred}}|$}{
    $B^{\text{pred}} \leftarrow \mathcal{B}^{\text{rem}}_{\text{pred}}[j]$\;
    $\mathcal{G}\leftarrow\emptyset$ \tcp*{matched GT blocks for this predicted block}

    \BlankLine
    \textbf{(2.1) Wrap match: predicted block covers GT block start.}\;
    \ForEach{$B^{\text{gt}}\in \mathcal{B}^{\text{rem}}_{\text{gt}}$}{
        \If{$B^{\text{pred}}.\text{start} \le B^{\text{gt}}.\text{start} \le B^{\text{pred}}.\text{end}$}{
            $\mathcal{G}\leftarrow \mathcal{G}\cup\{B^{\text{gt}}\}$\;
        }
    }

    \BlankLine
    \textbf{(2.2) Near match: context-line overlap before and after.}\;
    \If{$\mathcal{G}=\emptyset$}{
        $S^{-}_{\text{pred}}\leftarrow \textsc{ContextBefore}(C_b,B^{\text{pred}},\mathcal{B}^{\text{rem}}_{\text{pred}})$\;
        $S^{+}_{\text{pred}}\leftarrow \textsc{ContextAfter}(C_b,B^{\text{pred}},\mathcal{B}^{\text{rem}}_{\text{pred}})$\;
        \ForEach{$B^{\text{gt}}\in \mathcal{B}^{\text{rem}}_{\text{gt}}$}{
            $S^{-}_{\text{gt}}\leftarrow \textsc{StrideContextBefore}(C_b,B^{\text{gt}})$\;
            $S^{+}_{\text{gt}}\leftarrow \textsc{StrideContextAfter}(C_b,B^{\text{gt}})$\;
            \If{$\textsc{LineSetMatch}(S^{-}_{\text{pred}},S^{-}_{\text{gt}})$ \textbf{and}
                $\textsc{LineSetMatch}(S^{+}_{\text{pred}},S^{+}_{\text{gt}})$}{
                $\mathcal{G}\leftarrow \{B^{\text{gt}}\}$\;
                \textbf{break}\;
            }
        }
    }

    \BlankLine
    \textbf{(2.3) Distant-but-identical: edit-wise equality.}\;
    \If{$\mathcal{G}=\emptyset$}{
        \ForEach{$B^{\text{gt}}\in \mathcal{B}^{\text{rem}}_{\text{gt}}$}{
            \If{$|B^{\text{pred}}.\Delta|=|B^{\text{gt}}.\Delta|$ \textbf{and} $\textsc{Equal}(B^{\text{pred}}.\Delta.\text{edit}, B^{\text{gt}}.\Delta.\text{edit})$}{
                $\mathcal{G}\leftarrow \{B^{\text{gt}}\}$ \tcp*{same edits in the same order, at any location}
                \textbf{break}\;
            }
        }
    }

    \BlankLine
    \If{$\mathcal{G}\neq\emptyset$}{  
        $B^{\text{test}} \leftarrow \textsc{MergeBlocks}(\mathcal{B}^{\text{all}}_{\text{gt}} \setminus \gG \cup \{B^{\text{pred}}\})$\;
        $C^{\text{test}} \leftarrow \textsc{Apply}(B^{\text{test}}.\Delta, C_b)$\;
        $\mathcal{M}\leftarrow \mathcal{M}\cup\{\textsc{MakeMatch}(B^{\text{pred}},\mathcal{G}, C^{\text{test}})\}$ \tcp*{Use $C^{\text{test}}$ to test matched pairs}
        remove all blocks in $\mathcal{G}$ from $\mathcal{B}^{\text{rem}}_{\text{gt}}$\;
    }
}
\Return{$\mathcal{M}$}\;
\end{algorithm*}

\begin{algorithm*}
\caption{$\ess_\gU$: finding $\epsilon$-relaxed essential edits for each matching in $\gM$}
\label{alg:map_epsilon}
\DontPrintSemicolon
\KwIn{Buggy program $C_b$; GT blocks $\mathcal{B}_{\text{gt}}$; Predicted and GT edits $\Delta_{\text{pred}}, \Delta_{\text{gt}}$; Tolerance $\epsilon$; Unit tests $F_{\gU}(\cdot)$}
\KwOut{Final matching $\mathcal{M}_\epsilon$ with two more additional fields \texttt{success} and \texttt{essential\_size}}

\BlankLine
\textbf{Step 1: Candidate matching via $\map$.}\;
$\mathcal{M} \leftarrow \map(C_b,\Delta_{\text{pred}},\Delta_{\text{gt}})$\;
\tcp{$\gM$ contains a pred\_block $B_\text{pred}$, gt\_blocks $\gG$, and a tester $C^{\text{test}}$ built by replacing matched GT blocks with predicted blocks.}
$\kappa(\gG) \leftarrow \sum_{B^{\text{gt}}\in\gG}\big(|B^{\text{gt}}.\Delta|+\epsilon\big)$\;
\tcp{Allowed predicted lines for a match on GT blocks $\gG$: $|E_i|+\epsilon$ per matched bug $i$, i.e., $|\gG|\cdot(\epsilon+1)$ for single-line bugs}

\BlankLine
\textbf{Step 2: Semantic equivalence verification using $F_{\gU}$.}\;
\ForEach{match record $r \in \gM$ }{
    \eIf{$r.\texttt{tester} = \text{none}$ \textbf{or} $F_{\gU}(r.\texttt{tester})=1$}{
        $r.\texttt{success}\leftarrow \texttt{True}$\;
        $B^{\text{pred}} \leftarrow r.\texttt{pred\_block}$\; 
        $\gG \leftarrow r.\texttt{gt\_blocks}$\;
        $r.\texttt{essential\_size}\leftarrow \min{(\kappa(\gG), |B^{\text{pred}}.\Delta|)}$\;
    }{
        $r.\texttt{success}\leftarrow \texttt{False}$\;
        $r.\texttt{essential\_size}\leftarrow 0$\;
    }
}

\BlankLine
\textbf{Step 3: Deep redundancy check to realize $\epsilon$-relaxed essential edits.}\;
\ForEach{match record $r \in \mathcal{M}$ \textbf{with} $r.\texttt{success}=\texttt{True}$}{
    $\mathcal{S}\leftarrow\emptyset$ \tcp*{candidate sub-blocks}
    $B^{\text{pred}} \leftarrow r.\texttt{pred\_block}$\; 
    $\gG \leftarrow r.\texttt{gt\_blocks}$\;

    \BlankLine
    \tcp{Enumerate smaller contiguous sub-edits within the predicted block.}
    let $(B^{\text{pred}}.\Delta) = [(\ell_1,v_1),\dots,(\ell_m,v_m)]$ ordered by $\ell$\;
    \For{$\tau \leftarrow 0$ \KwTo $\kappa(\gG) - 1$}{
        \For{$line \leftarrow 1$ \KwTo $m-\tau$}{
            $B^{\text{sub}} \leftarrow \textsc{SubBlock}(B^{\text{pred}}, line, line + \tau)$\;
            $B^{\text{test}} \leftarrow \textsc{MergeBlocks}(\mathcal{B}_{\text{gt}} \setminus \gG \cup \{B^{\text{sub}}\}$)\;
            $C^{\text{sub}} \leftarrow \textsc{Apply}(B^{\text{test}}.\Delta, C_b)$\;
            $\mathcal{S} \leftarrow \mathcal{S}\cup\{(\tau + 1,C^{\text{sub}})\}$\;
        }
    }

    \BlankLine
    \tcp{Find the smallest $\tau$ that still passes $F_{\gU}$.}
    $\tau^\star \leftarrow +\infty$\;
    \ForEach{$(\tau,C^{\text{sub}})\in \mathcal{S}$}{
        \If{$F_{\gU}(C^{\text{sub}})=1$ \textbf{and} $\tau < \tau^\star$}{
            $\tau^\star \leftarrow \tau$\;
        }
    }
    \If{$\tau^\star < +\infty$}{
        $r.\texttt{essential\_size}\leftarrow \tau^\star$\;
    }
}

\BlankLine
$\mathcal{M}_\epsilon \leftarrow \mathcal{M}$\;
\Return{$\mathcal{M}_\epsilon$}\;
\end{algorithm*}

\clearpage

\section{Examples of Debugging Categories}
\label{appsec:debugcate}

We supply the qualitative examples behind the failure-mode taxonomy reported in \S\ref{sec:metric_verf}. We separate examples into two regimes that the unit-test signal alone cannot distinguish: (i) patches that pass the unit tests but receive imperfect precision or recall (Figures~\ref{fig:redundant-guard}--\ref{fig:bug-composition}), and (ii) patches that fail the unit tests despite containing partially correct edits (Figures~\ref{fig:under-repair}--\ref{fig:new-bug}). Percentages cited in each caption are computed over the 240 manually inspected patches drawn for the metric-verification study.

\subsection{Failure modes when unit tests pass}

\begin{figure*}[ht]
    \centering
    \includegraphics[width=\linewidth]{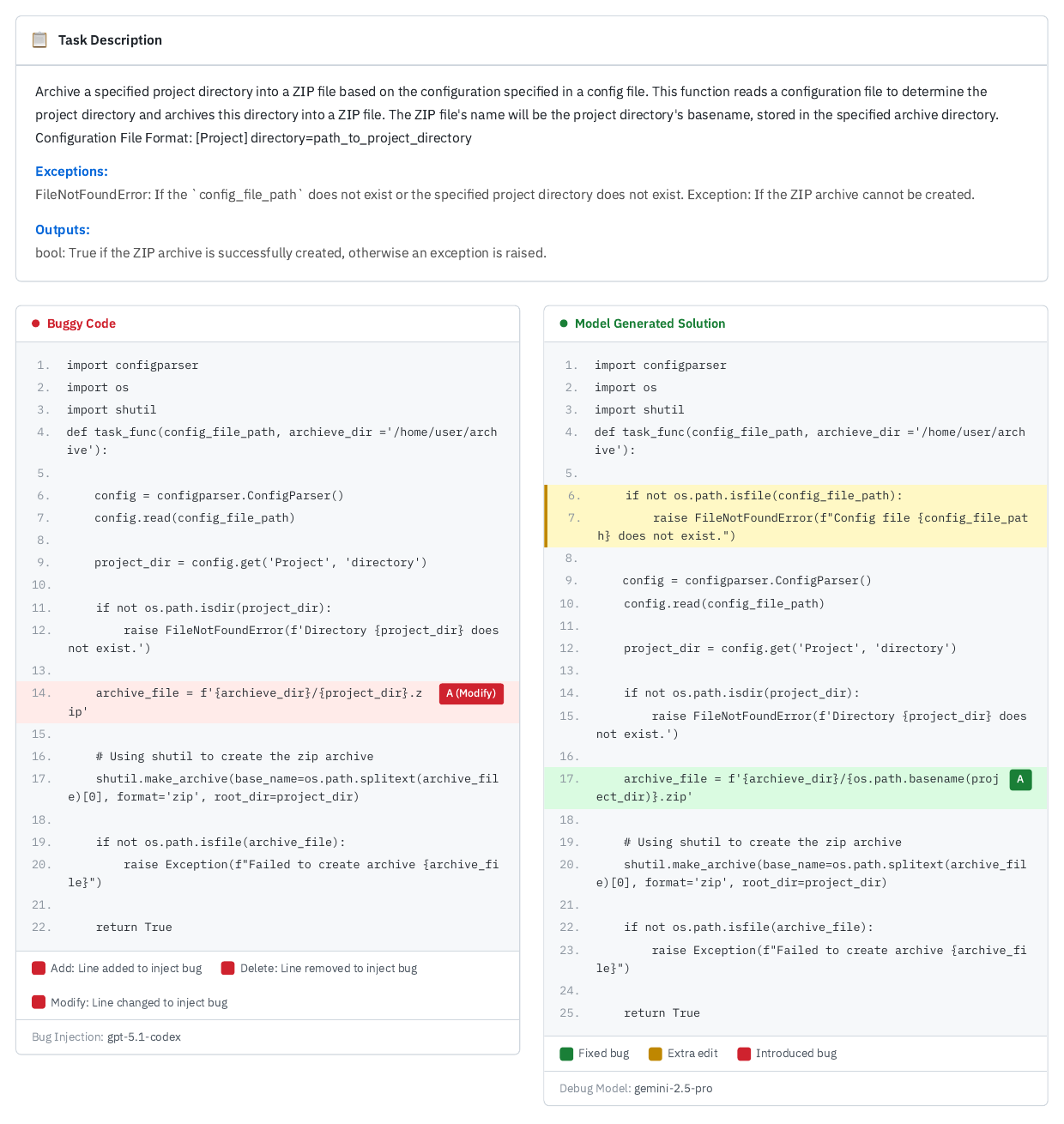}
    \caption{Redundant guard checks (9.8\% of recall=1, precision$<$1 cases). The model adds a defensive check (e.g., a null guard) on top of an already-correct fix; the extra check is irrelevant to the unit tests but counts as a spurious edit.}
    \label{fig:redundant-guard}
\end{figure*}

\begin{figure*}[ht]
    \centering
    \includegraphics[width=\linewidth]{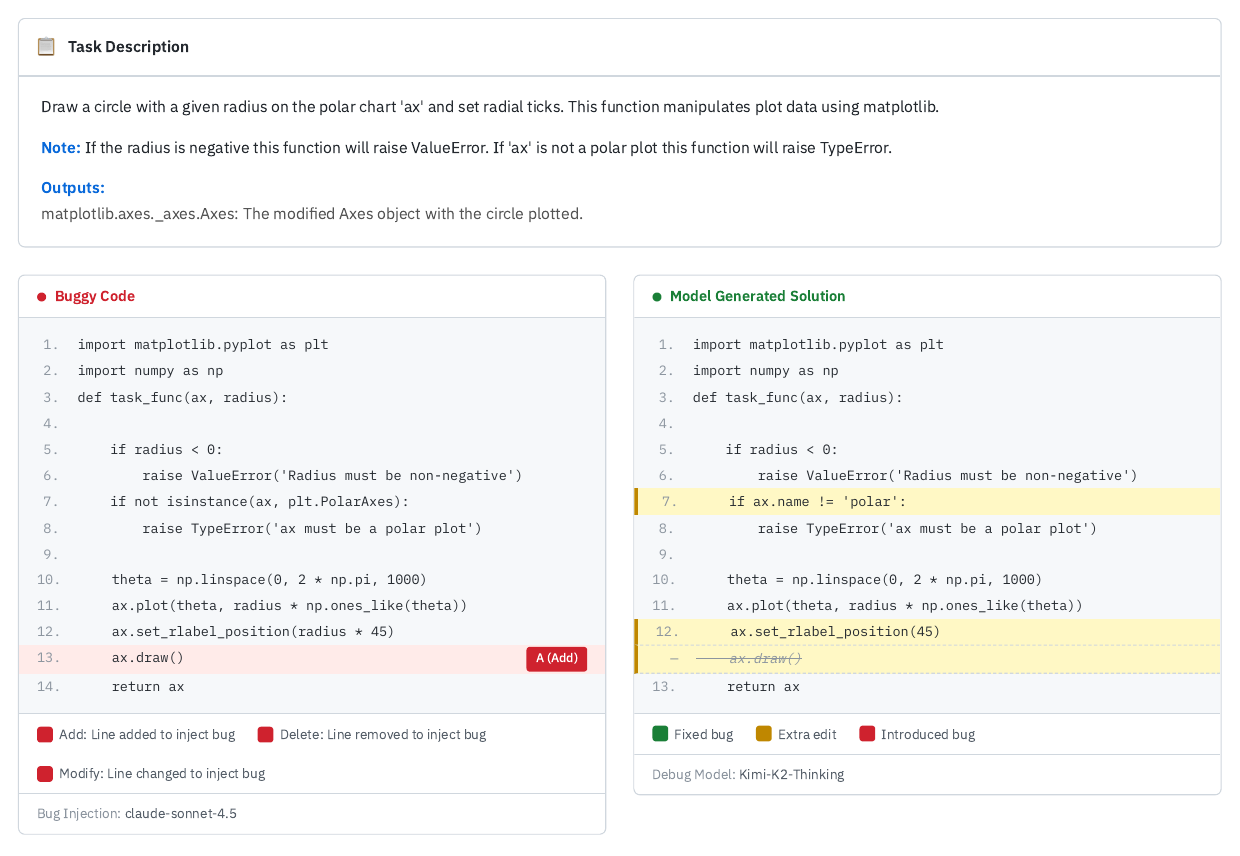}
    \caption{Additional modifications to correct code (66.8\% of recall=1, precision$<$1 cases). The model edits unrelated, already-correct lines, typically renaming variables, reordering statements, or adjusting unrelated control flow, in addition to applying the necessary fix.}
    \label{fig:extra-modification}
\end{figure*}

\begin{figure*}[ht]
    \centering
    \includegraphics[width=\linewidth]{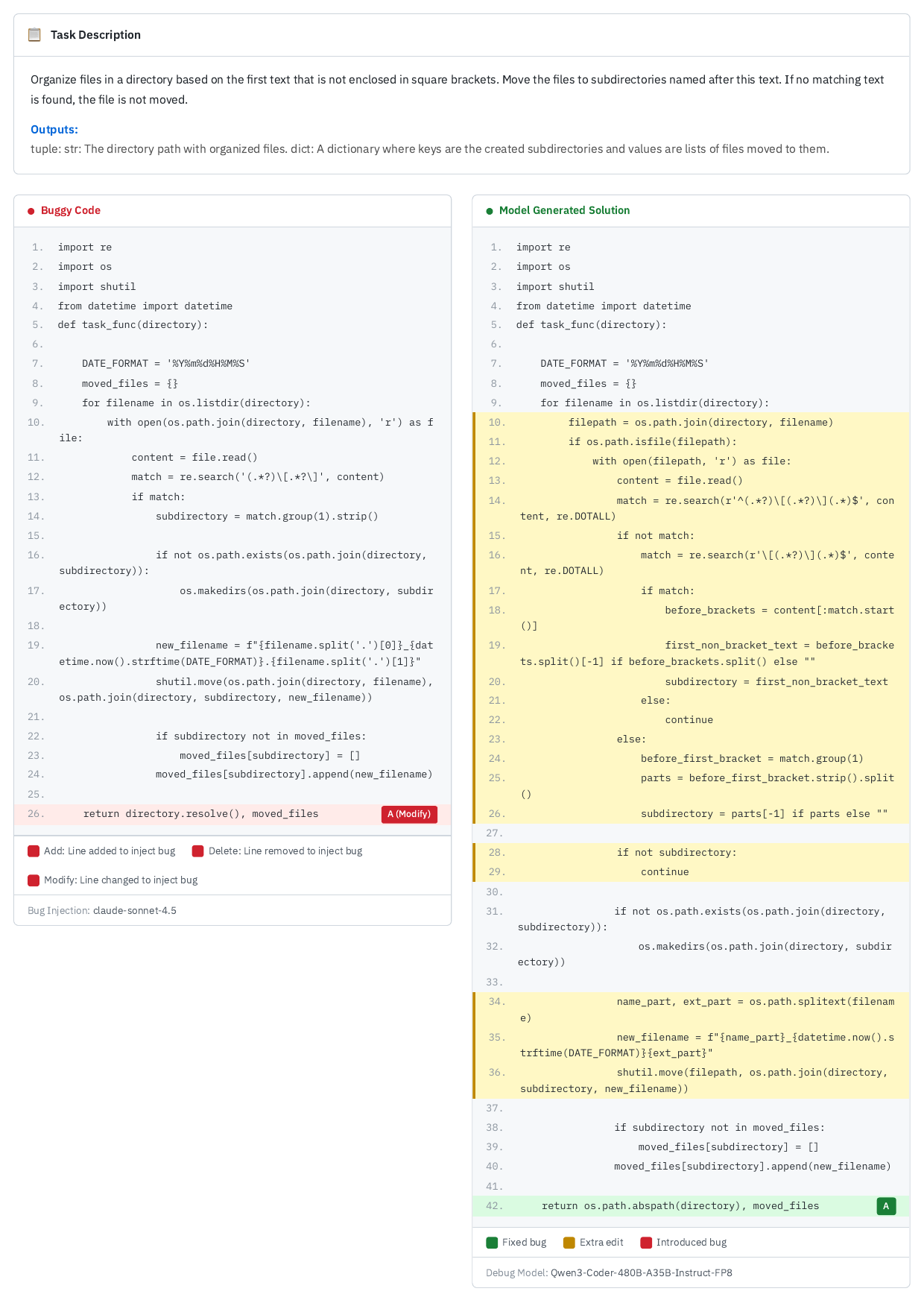}
    \caption{Complete rewrite (7.8\% of recall=1, precision$<$1 cases). The model regenerates the entire function from scratch; the original buggy lines are absent from the patch, replaced by an alternative implementation.}
    \label{fig:complete-rewrite}
\end{figure*}

\begin{figure*}[ht]
    \centering
    \includegraphics[width=\linewidth]{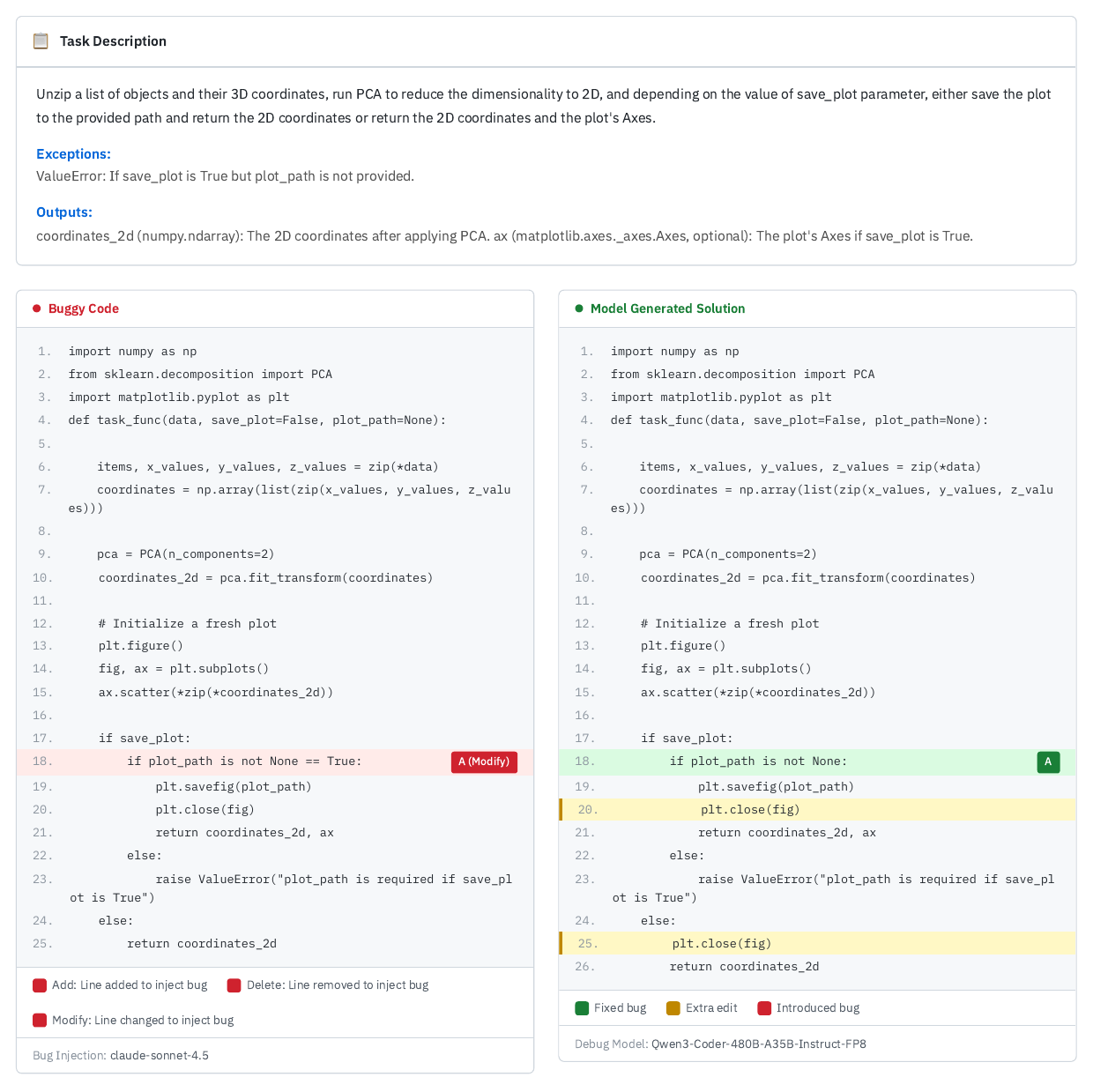}
    \caption{Discovering bugs missed by ground truth (1.9\% of recall=1, precision$<$1 cases). The model fixes the injected bug \emph{and} a separate, latent defect in the seed solution that the original benchmark's tests fail to exercise.}
    \label{fig:gt-oversight}
\end{figure*}

\begin{figure*}[ht]
    \centering
    \includegraphics[width=\linewidth]{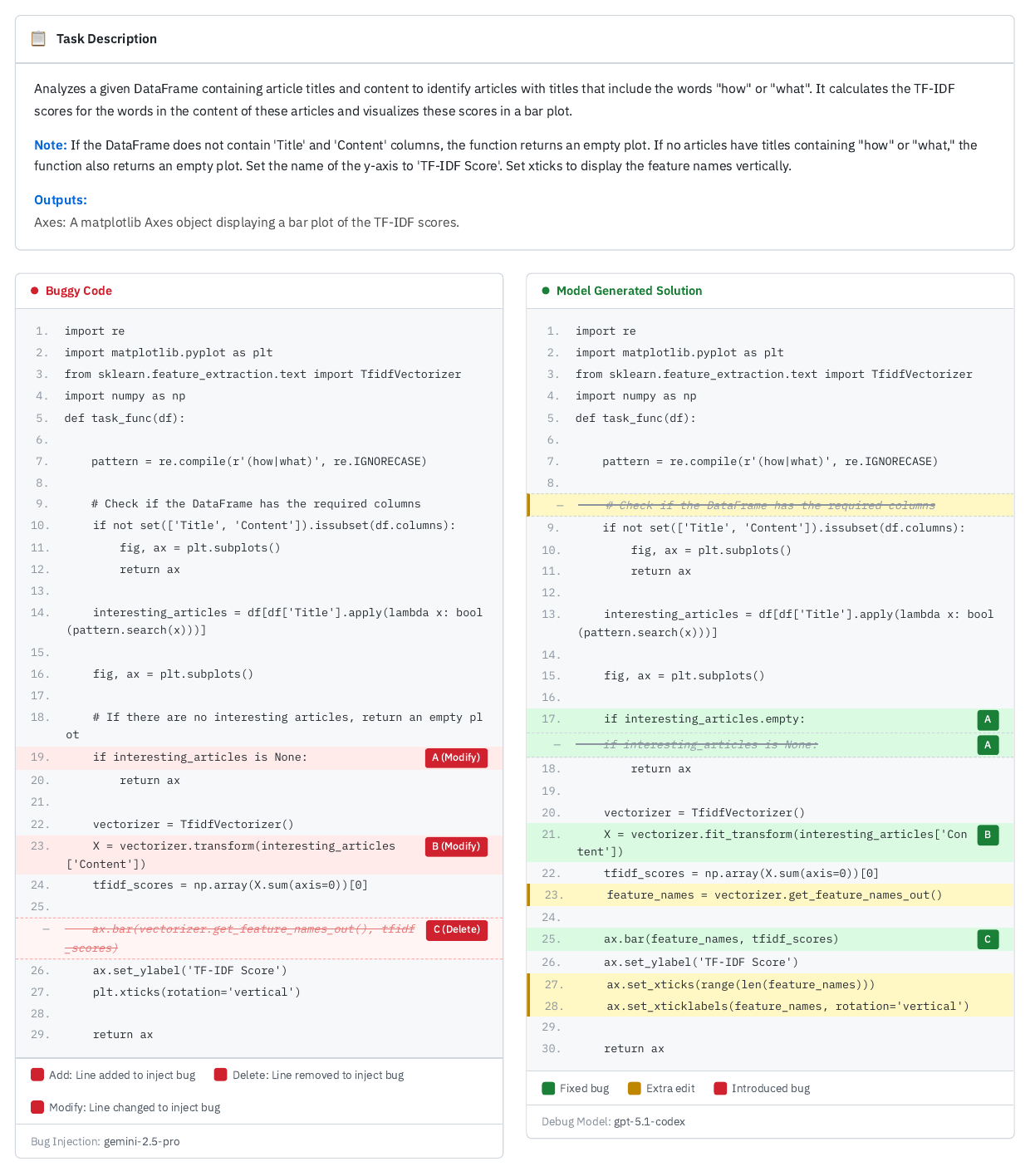}
    \caption{Functionally correct but undetected (70\% of recall$<$1 cases). The model's edit is semantically equivalent to the ground-truth fix but expressed at a different location or in a different syntactic form, so the strict block matcher fails to align it.}
    \label{fig:func-correct-undetected}
\end{figure*}

\begin{figure*}[ht]
    \centering
    \includegraphics[width=\linewidth]{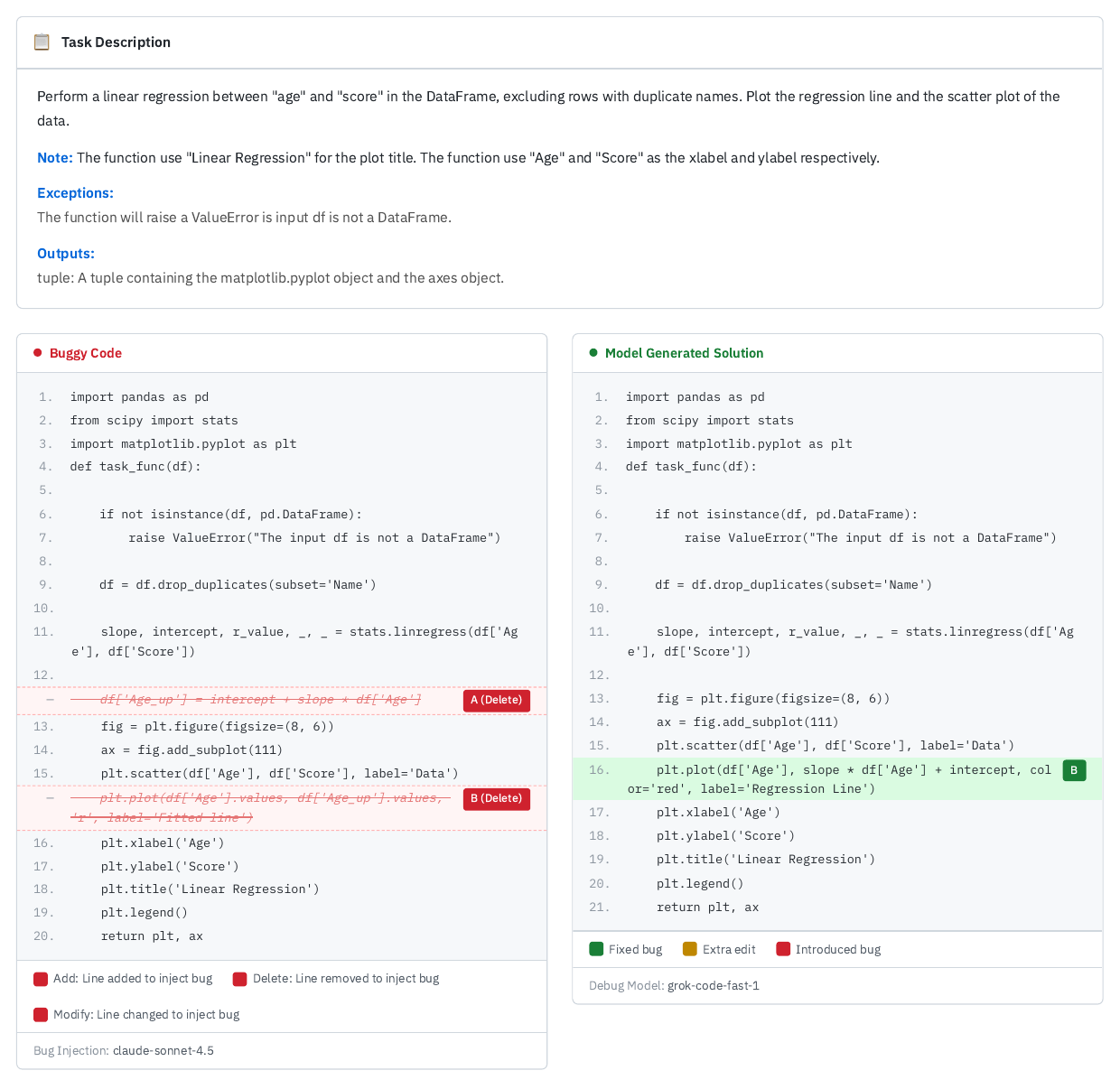}
    \caption{Multiple minimal fixes (20\% of recall$<$1 cases). The injected bug admits more than one minimal correct repair, and the model selects a fix that disagrees with the ground-truth choice while remaining minimal.}
    \label{fig:multiple-fixes}
\end{figure*}

\begin{figure*}[ht]
    \centering
    \includegraphics[width=\linewidth]{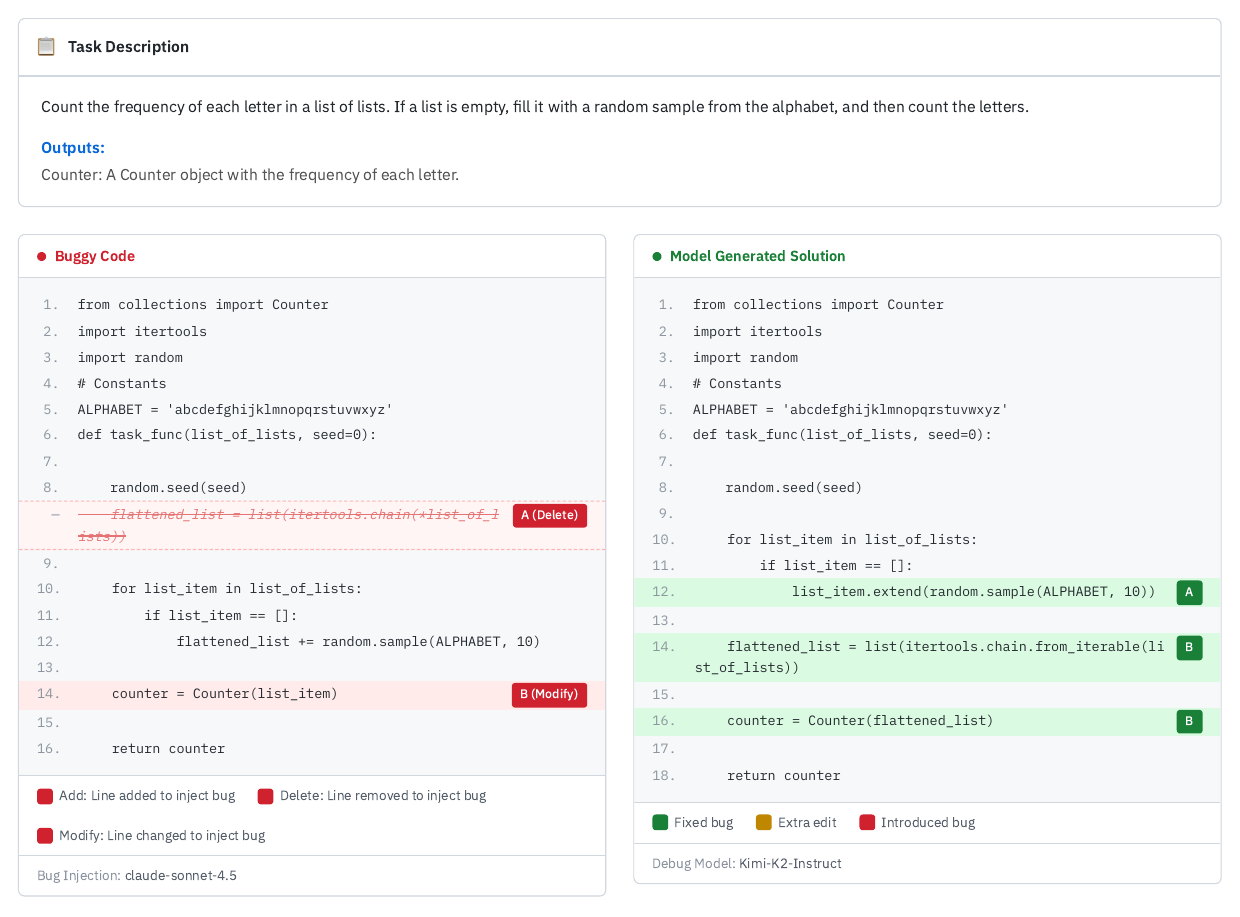}
    \caption{Bug-composition artefacts (10\% of recall$<$1 cases, $\approx$1.65\% overall). When multiple bugs are injected into the same program, an upstream injected bug occasionally changes the program's logic enough that a downstream injected bug becomes vacuous or fuses with the upstream fix.}
    \label{fig:bug-composition}
\end{figure*}

The seven figures below stratify cases where the unit tests pass but the patch is not maximally targeted. Figures~\ref{fig:redundant-guard}--\ref{fig:gt-oversight} cover precision-loss patterns (recall $=1$, precision $<1$); Figures~\ref{fig:func-correct-undetected}--\ref{fig:bug-composition} cover recall-loss patterns (recall $<1$, where the recovered fix is functionally equivalent or affected by injection coupling).

Figure~\ref{fig:redundant-guard} illustrates the most benign precision loss: the model produces the correct minimal fix and then layers an additional guard on top. The guard is harmless, often even reasonable defensive practice, but it is a strict superset of the ground-truth edit and therefore reduces precision without affecting unit-test outcome.

Figure~\ref{fig:extra-modification} captures the dominant precision-loss mode: while the necessary fix is present, the patch also rewrites surrounding correct code. This is the empirical signature of the regeneration-style behaviour discussed in \S\ref{sec:results}; the model treats the buggy program as a draft to revise globally rather than a working program to amend locally.

Figure~\ref{fig:complete-rewrite} isolates the most extreme precision failure: the patch is a from-scratch reimplementation rather than an edit. Such patches almost always pass the unit tests by construction (the rewrite is correct) but score precision near zero because no predicted edit aligns with a ground-truth block; the model has effectively bypassed the debugging task.

Figure~\ref{fig:gt-oversight} highlights a category in which low precision is genuinely informative: the model patches an additional, real bug in the seed solution itself. These cases account for under 2\% of the inspected patches but illustrate that perfect precision is not always desirable when the ground truth is incomplete; we interpret them as a soft upper bound on how aggressively precision can be optimized without losing useful repairs.

Figure~\ref{fig:func-correct-undetected} dominates the recall-loss regime. The fix is correct on the test suite, but the predicted edit is structurally far enough from the ground-truth block that $\map$ cannot match them. These cases are the target of the relocation-tolerant criteria in Pass~2 of $\map$ (near match and distant-but-identical; Appendix~\ref{appsec:algo}), which pair moved edits with their ground-truth blocks and validate them with unit tests; edits that are both moved and rewritten remain beyond their reach.

Figure~\ref{fig:multiple-fixes} captures the irreducible ambiguity that comes from defining recall against a single reference patch. When two or more equally targeted fixes resolve the same bug, the model is penalized for choosing the wrong one even though no over-editing occurred. The 20\% prevalence among recall-loss cases is consistent with bug-pattern diversity in the ODC categories; addressing it would require multi-reference annotations, which we discuss as a future direction in \S\ref{sec:results}.

Figure~\ref{fig:bug-composition} is the only recall-loss category attributable to dataset construction rather than model behaviour. Composing multiple bug injections risks turning two locally minimal injections into one coupled defect; the model then resolves the pair with a single edit that the matcher counts as missing one of the original bugs. The 1.65\% overall prevalence indicates that the bug-independence assumption used by our generator holds in the large majority of cases, and that the metric remains accurate on $>$97.5\% of the evaluation data.

\subsection{Failure modes when unit tests fail}

The two figures below cover the symmetric regime: patches that fail the unit tests despite containing partially correct edits. The taxonomy here is binary, separating cases where the model under-edits from cases where it edits enough but introduces fresh defects.

\begin{figure*}[ht]
    \centering
    \includegraphics[width=\linewidth]{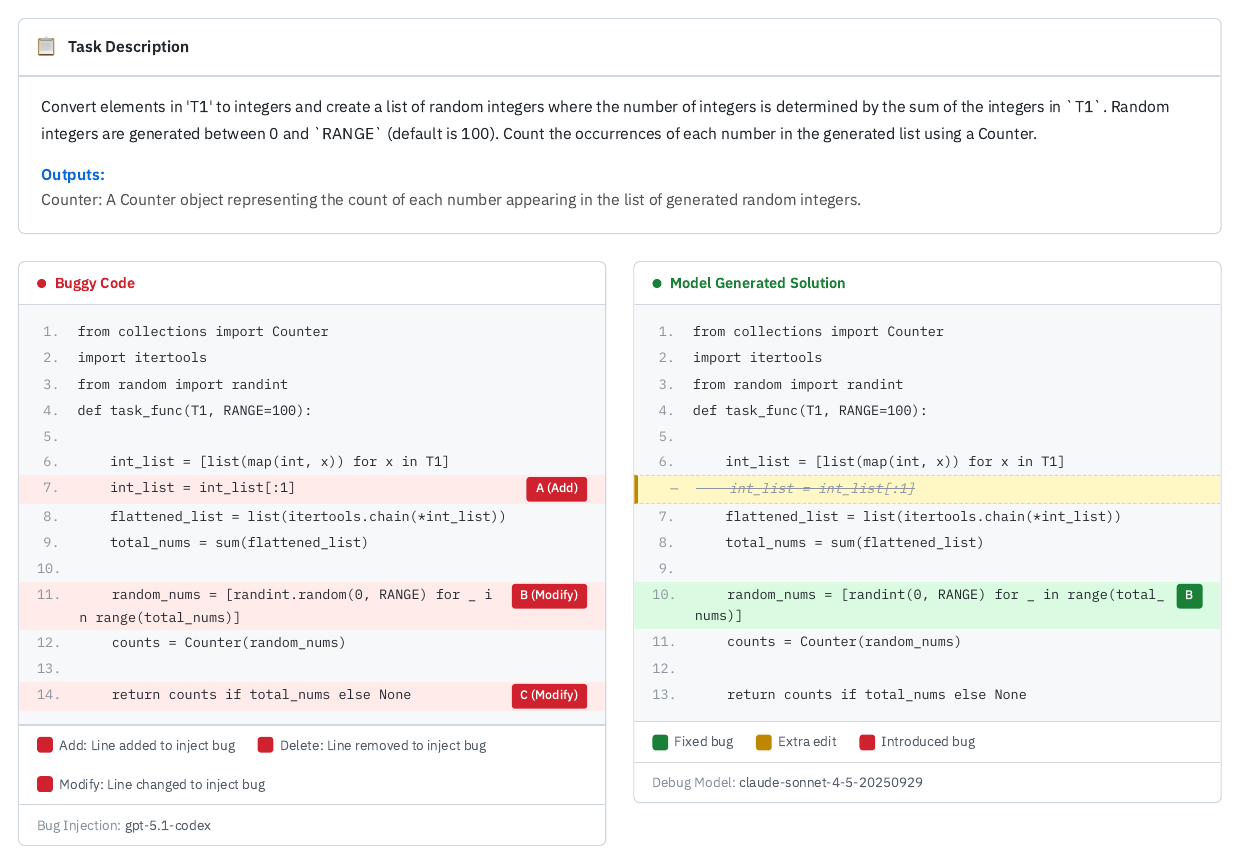}
    \caption{Under-repair (31.4\% of failing cases). The model applies a subset of the necessary fixes without introducing extraneous edits (precision $=1$), but at least one ground-truth bug is left untouched (recall $<1$), so the unit tests fail.}
    \label{fig:under-repair}
\end{figure*}

Figure~\ref{fig:under-repair} shows a category that, despite failing tests, exhibits exactly the targeted-edit behaviour the benchmark rewards: the model identifies and fixes some bugs precisely. Failures here are recall-bound rather than precision-bound, and they tend to grow with bug count $k$, consistent with the per-bug-count breakdown in Figure~\ref{fig:bug-count-breakdown-all}.

\begin{figure*}[ht]
    \centering
    \includegraphics[width=\linewidth]{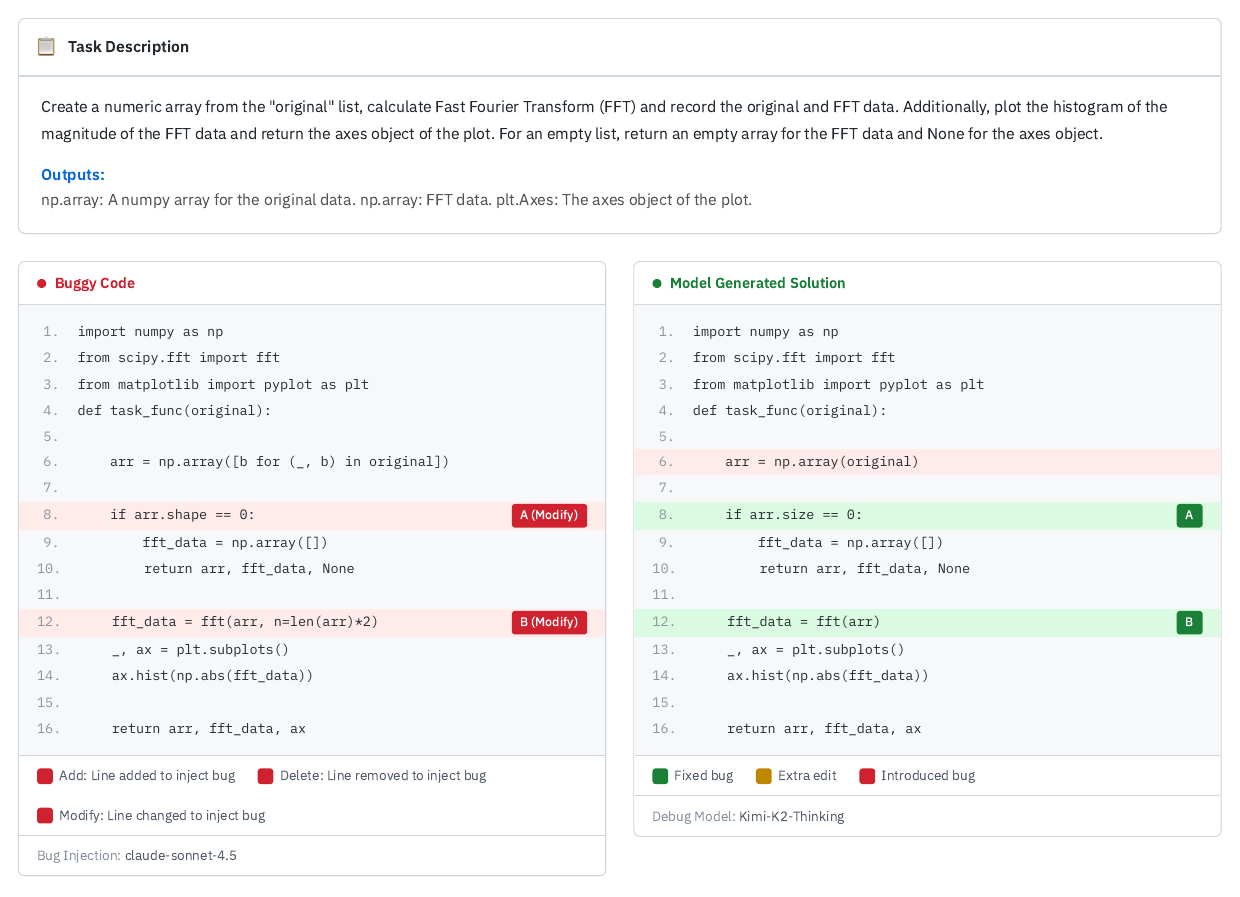}
    \caption{Regressive repair (39.2\% of failing cases). The model successfully resolves all originally injected bugs (recall $=1$) but introduces new defects elsewhere in the patch that cause the unit tests to fail.}
    \label{fig:new-bug}
\end{figure*}

Figure~\ref{fig:new-bug} reveals the largest single failure category in this regime. The patch hits every ground-truth fix---recall is $1$---yet the unit tests still fail, because the model also rewrote correct code and broke it. This is the same regeneration mechanism that drives the precision losses in Figure~\ref{fig:extra-modification}; the difference is that the side effects landed in code paths the test suite happens to exercise. Together with under-repair, regressive repair shows that the unit-test pass rate alone discards information that precision and recall recover.

\clearpage

\section{Prompt templates}
\label{appsec:prompt}
\begin{figure*}[ht]
    \centering
    \includegraphics[width=\linewidth]{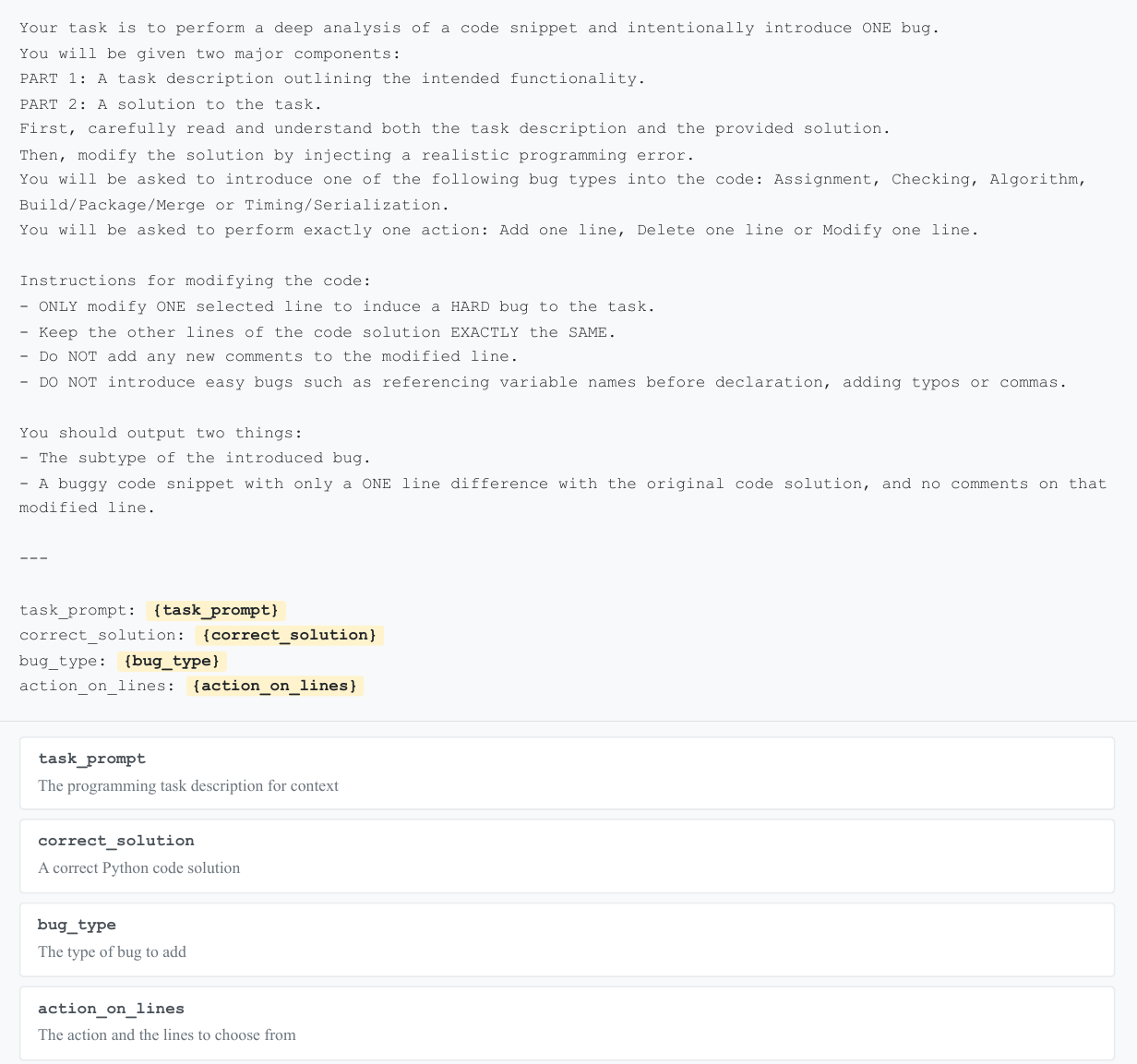}
    \caption{Bug injection prompt for benchmark construction.}
    \label{fig:prompt-bug-inject}
\end{figure*}

\begin{figure*}[ht]
    \centering
    \includegraphics[width=\linewidth]{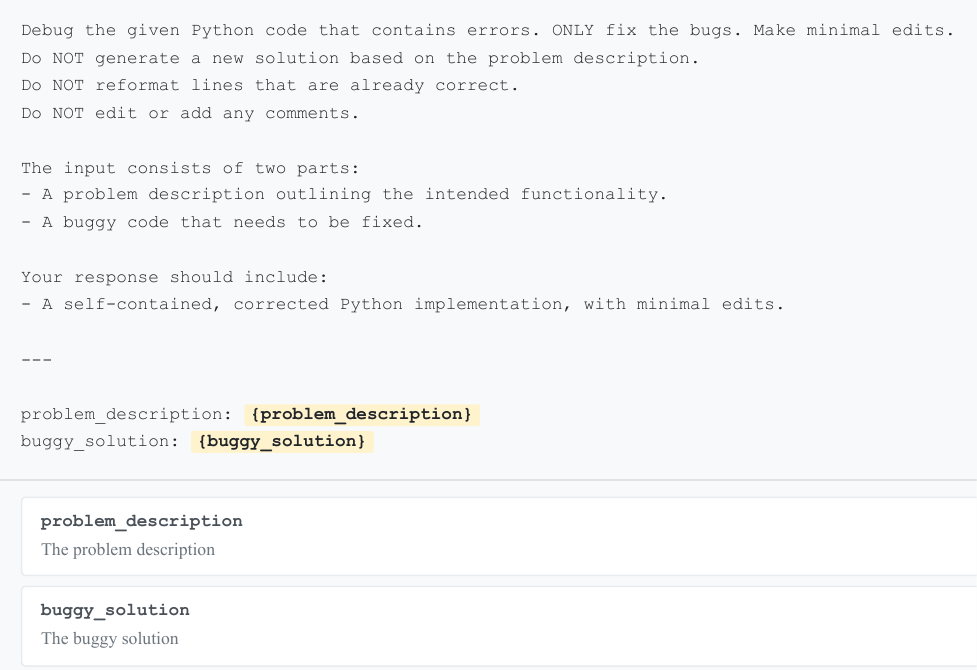}
    \caption{Minimal debugging prompt with problem description and buggy code.}
    \label{fig:prompt-minimal-debug}
\end{figure*}

\begin{figure*}[ht]
    \centering
    \includegraphics[width=\linewidth]{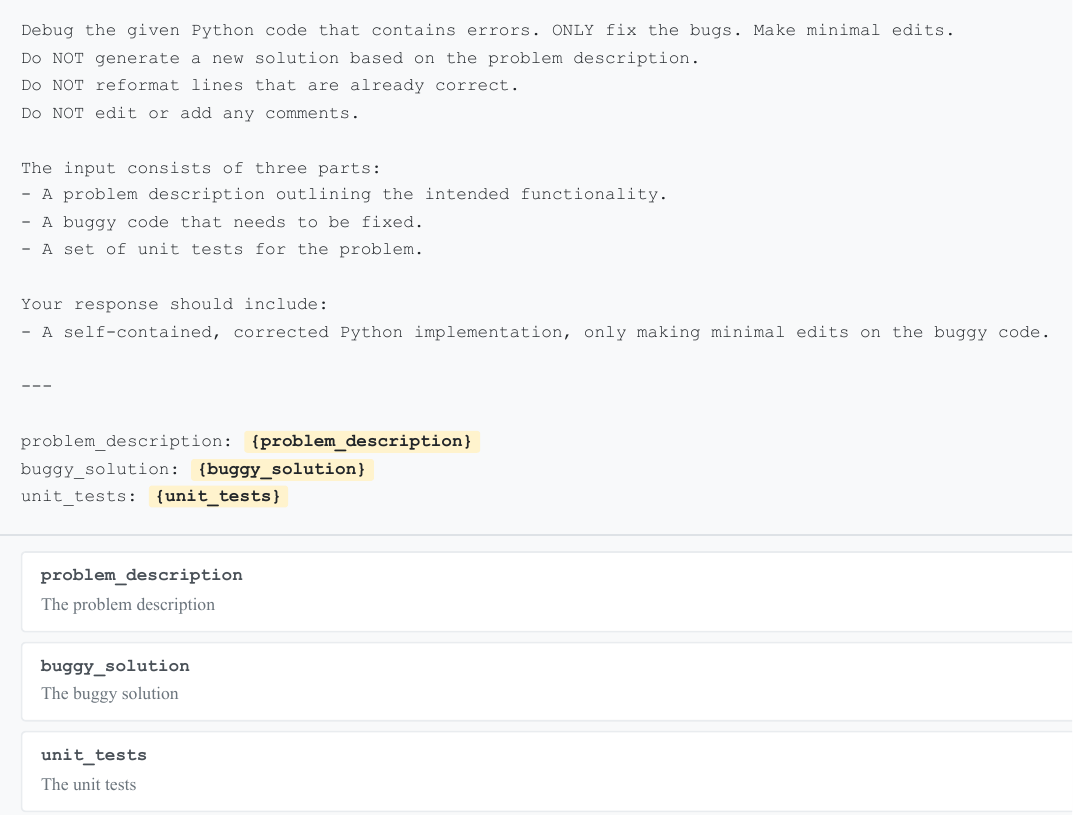}
    \caption{Minimal debugging prompt with unit tests.}
    \label{fig:prompt-minimal-unit}
\end{figure*}

\begin{figure*}[ht]
    \centering
    \includegraphics[width=\linewidth]{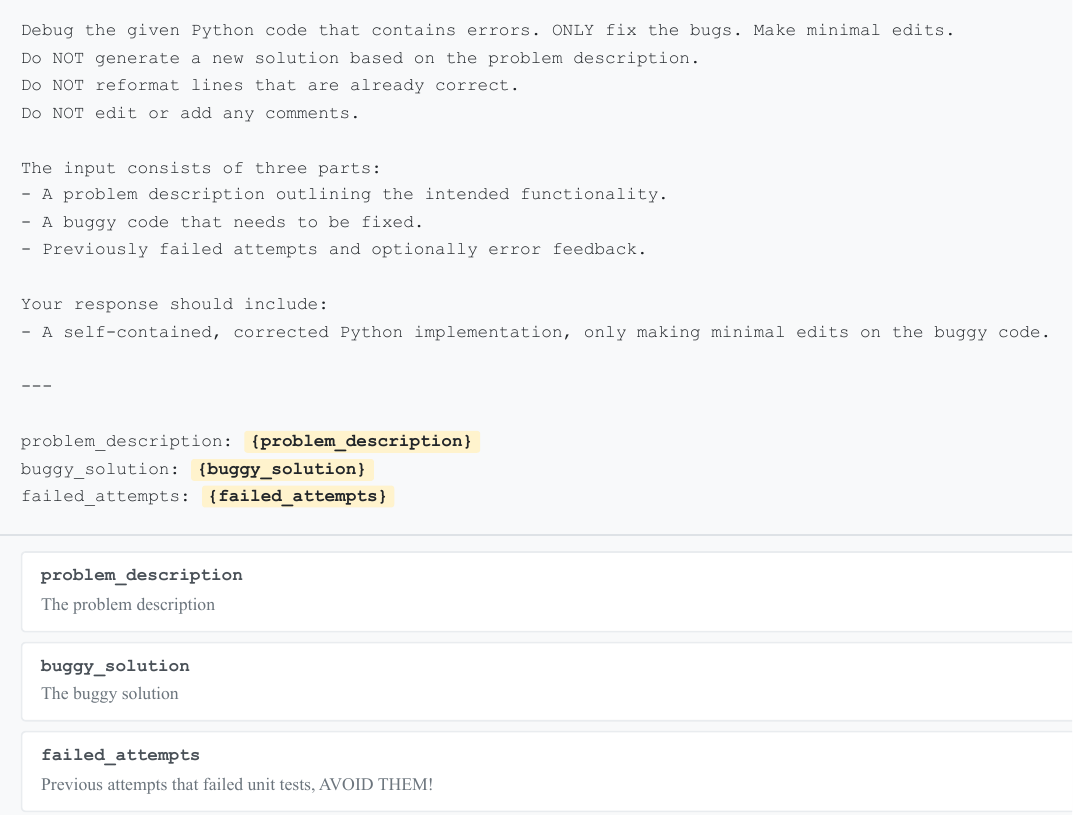}
    \caption{Minimal debugging prompt with execution feedback.}
    \label{fig:prompt-minimal-feedback}
\end{figure*}

\begin{figure*}[ht]
    \centering
    \includegraphics[width=\linewidth]{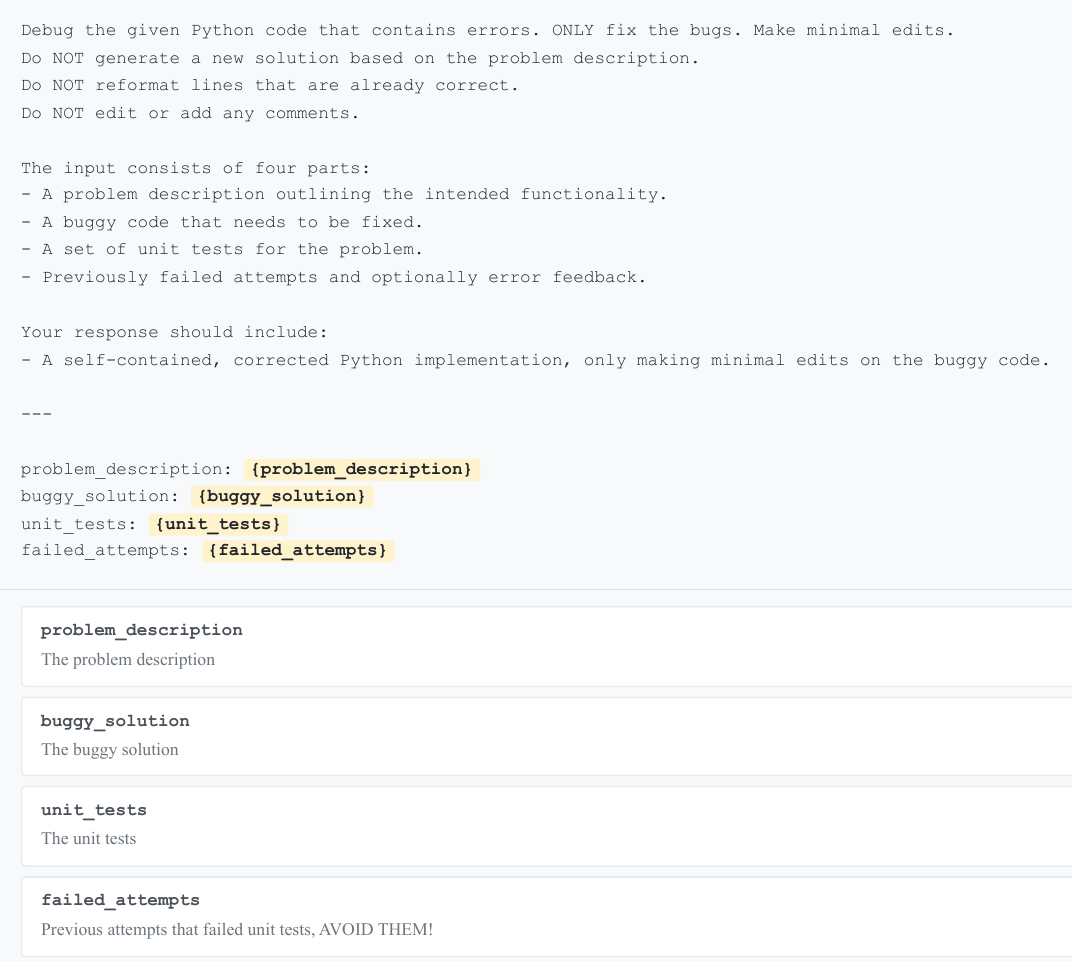}
    \caption{Minimal debugging prompt with unit tests and execution feedback.}
    \label{fig:prompt-minimal-unit-feedback}
\end{figure*}

\begin{figure*}[ht]
    \centering
    \includegraphics[width=\linewidth]{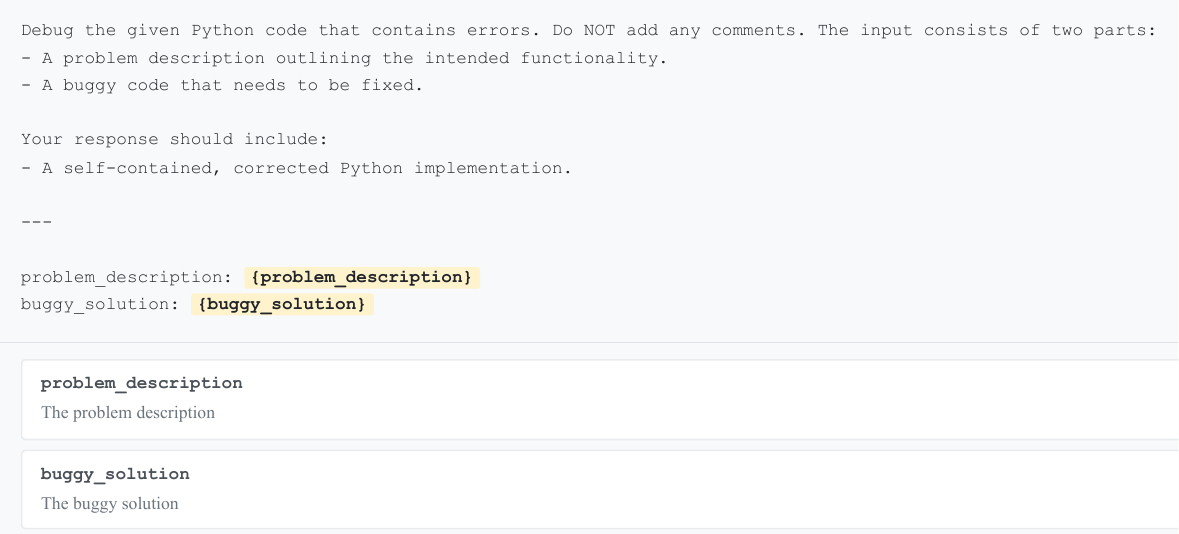}
    \caption{Free-form debugging prompt without minimal edit constraint.}
    \label{fig:prompt-free-debug}
\end{figure*}

\begin{figure*}[ht]
    \centering
    \includegraphics[width=\linewidth]{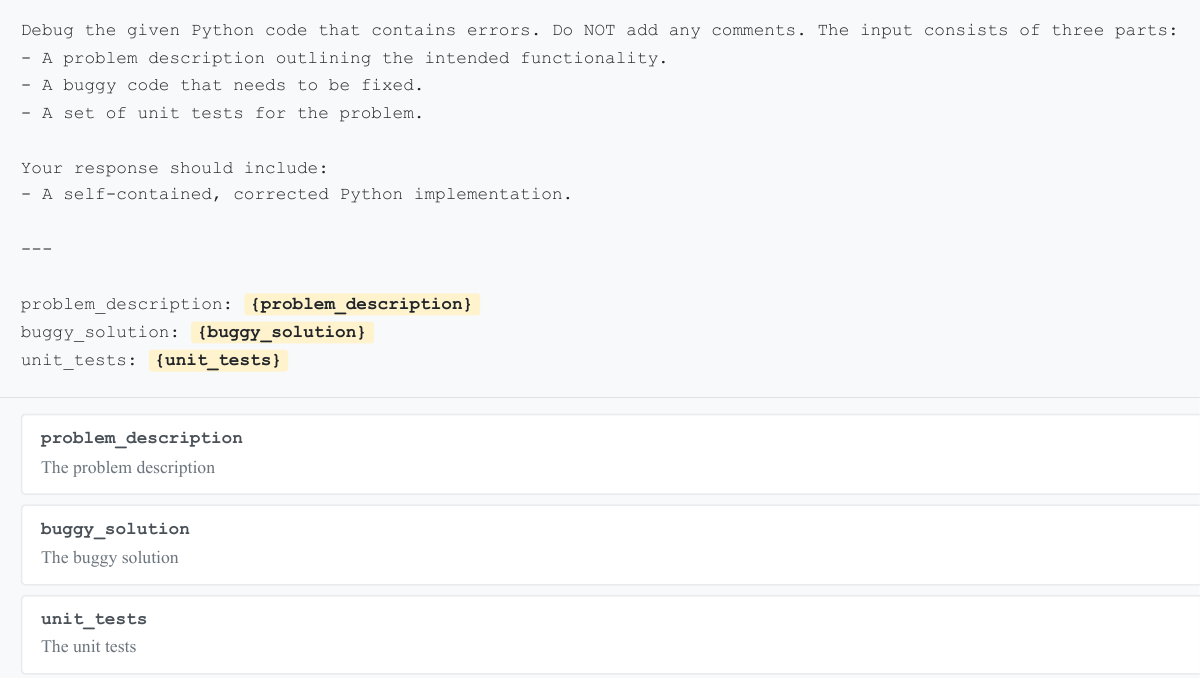}
    \caption{Free-form debugging prompt with unit tests.}
    \label{fig:prompt-free-unit}
\end{figure*}

\begin{figure*}[ht]
    \centering
    \includegraphics[width=\linewidth]{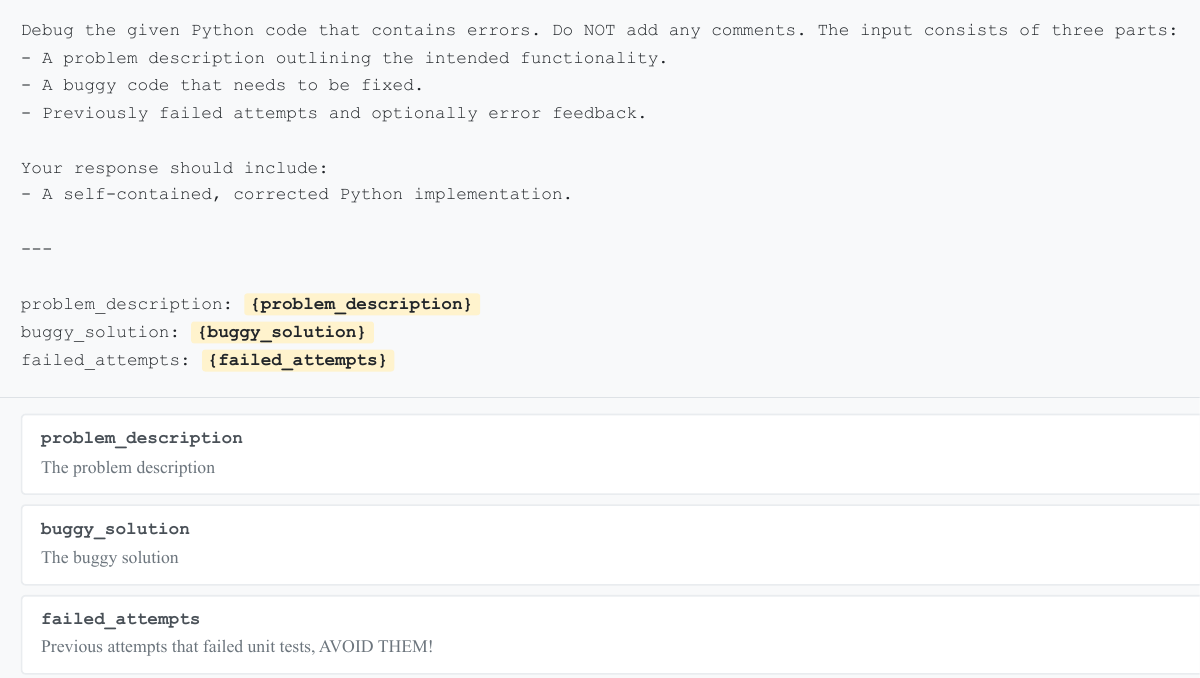}
    \caption{Free-form debugging prompt with execution feedback.}
    \label{fig:prompt-free-feedback}
\end{figure*}

\begin{figure*}[ht]
    \centering
    \includegraphics[width=\linewidth]{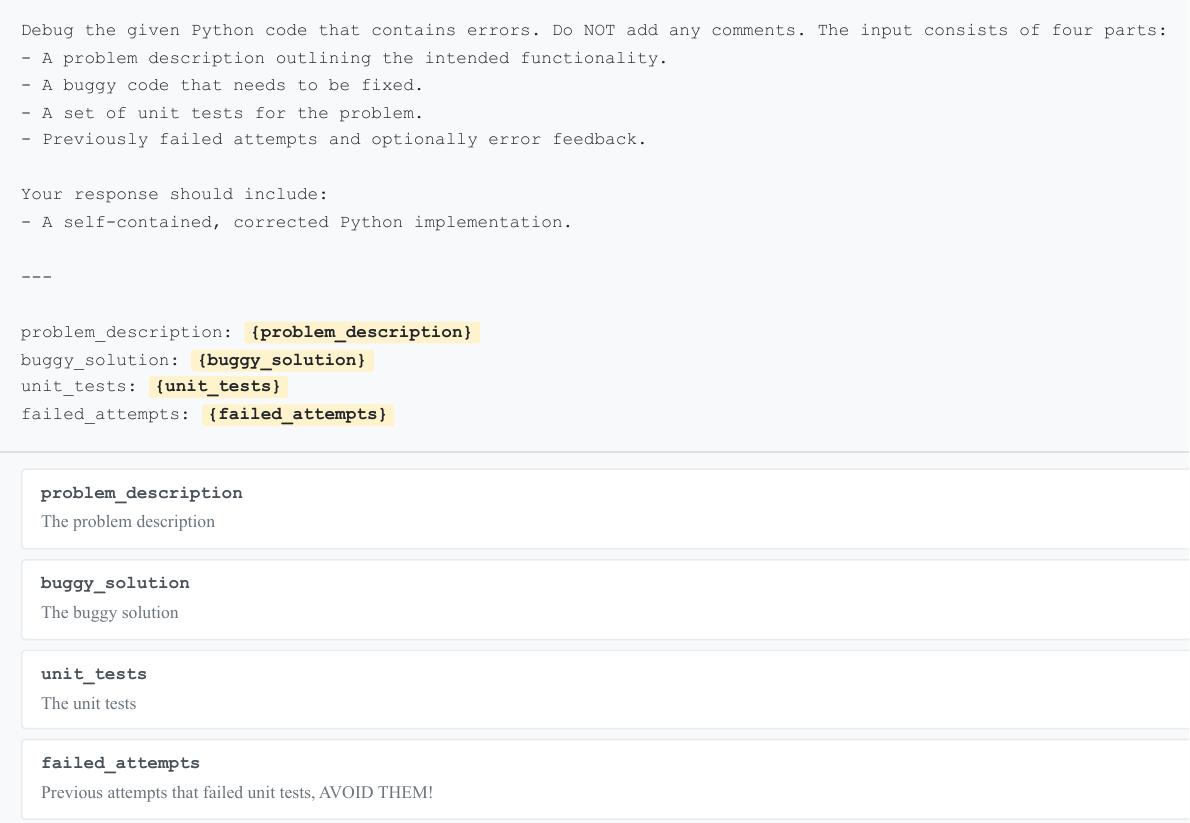}
    \caption{Free-form debugging prompt with unit tests and execution feedback.}
    \label{fig:prompt-free-unit-feedback}
\end{figure*}

\begin{figure*}[ht]
    \centering
    \includegraphics[width=\linewidth]{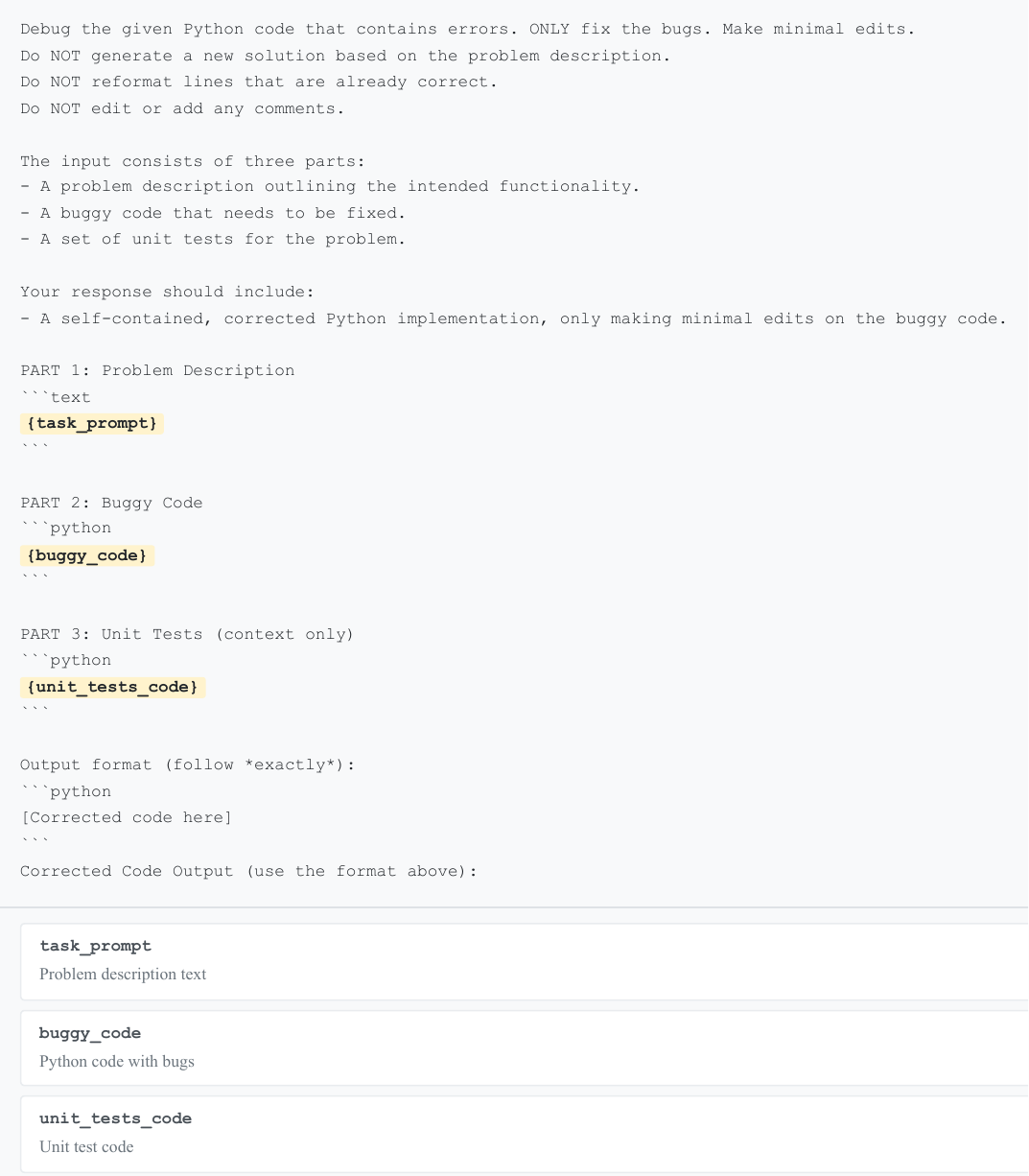}
    \caption{External API template for minimal debugging.}
    \label{fig:prompt-ext-minimal}
\end{figure*}

\begin{figure*}[ht]
    \centering
    \includegraphics[width=\linewidth]{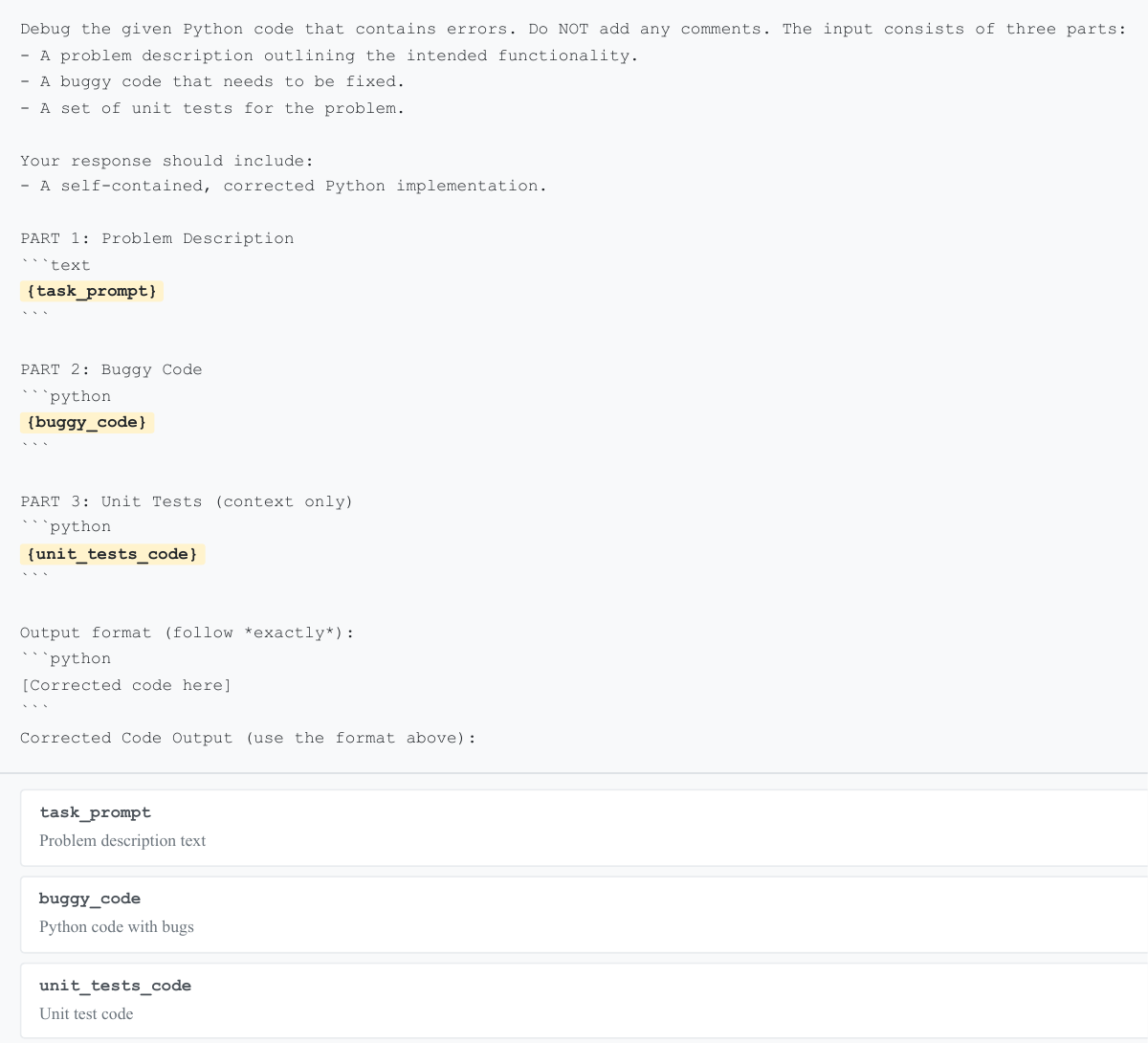}
    \caption{External API template for free-form debugging.}
    \label{fig:prompt-ext-free}
\end{figure*}

\begin{figure*}[ht]
    \centering
    \includegraphics[width=\linewidth]{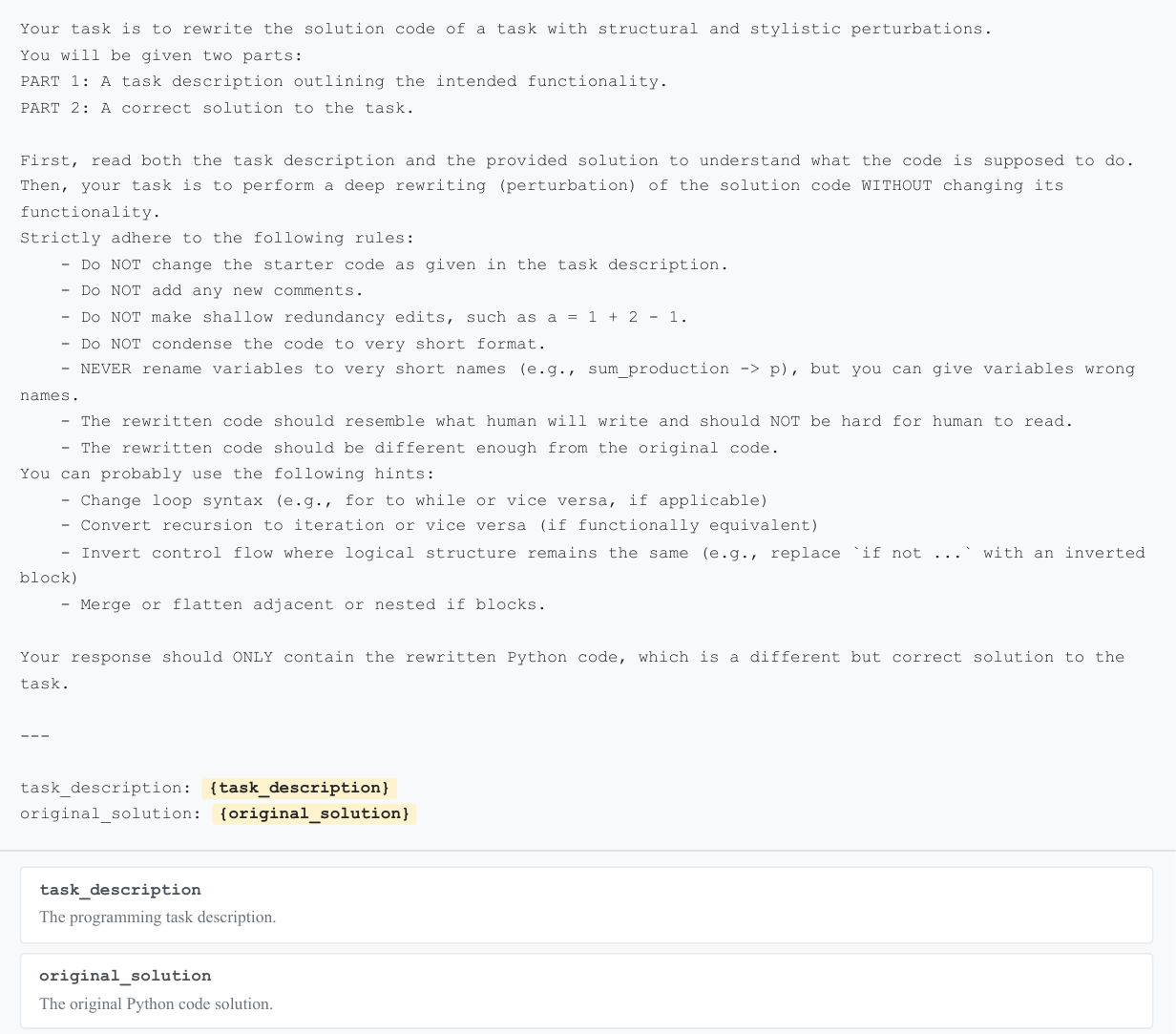}
    \caption{Solution rewriting prompt for benchmark construction.}
    \label{fig:prompt-rewrite}
\end{figure*}

We provide the prompt templates used in our experiments, ranging from bug injection and solution rewriting to minimal and free-form debugging with optional unit tests and execution feedback, shown in Figures~\ref{fig:prompt-bug-inject}–\ref{fig:prompt-rewrite}.

\clearpage

\section{Checklist Information}
\label{appsec:checklist}

\paragraph{Risks of malicious use of PDB pipeline.}
\pdb provides a systematic procedure for producing realistic buggy programs from existing code by prompting deliberate fault introduction. Hence, the same pipeline that supports controlled debugging evaluation could be repurposed for malicious bug-injection at scale, enabling automated generation of large quantities of plausible faulty code with minimal surface changes. Such capability may be misused to degrade software reliability in collaborative development settings, increase the review burden on maintainers, or seed low-quality code into shared repositories.

Another risk concern is potential data poisoning and model capability shaping. Because \pdb converts coding data into structured buggy program and solution pairs, it can lower the cost of creating large synthetic corpora that contain intentional buggy programs with minimal-edit transformations. If used outside the intended research context, these data could be silently employed to bias training toward behaviors that facilitate code degradation, or to contaminate downstream datasets used for model deployment and benchmarking. Even when the immediate artifacts are non-sensitive, the potential, silent shift in model behavior raises concerns. 


\paragraph{Licensing landscape of evaluated models.}
The governance of the evaluated LLMs and benchmarks reveals a sharp dichotomy between proprietary services and open-weight ecosystems. The proprietary tier includes GPT-5.1-Codex (OpenAI), Claude-Sonnet-4.5 (Anthropic), Gemini-2.5-Pro (Google DeepMind), and Grok-Code-Fast (xAI), all of which are accessible exclusively via commercial APIs. These systems are governed by restrictive Terms of Service that prohibit model weight extraction, reverse engineering, and competitive distillation, serving to protect their respective architectural innovations and agentic harnesses. 
In contrast, the open-weight landscape is characterized by permissive licensing designed to commoditize reasoning capabilities: DeepSeek-V3.2 and its reasoning variant DeepSeek-V3.2-Thinking are released under the MIT License, while Qwen3-Coder-480B utilizes the Apache License 2.0, which includes an explicit patent grant. 
A hybrid governance model is observed in Kimi-K2-Thinking and Kimi-K2-Instruct, which operate under a Modified MIT License; this variant permits general commercial use but mandates strictly visible attribution for entities exceeding 100 million monthly active users or \$20 million in monthly revenue. 

\paragraph{Licensing landscape of datasets.}
Regarding evaluation frameworks, \bigcode is governed by the Apache License 2.0, whereas \livecode adopts a split licensing model with its codebase under the MIT License and its dataset artifacts available under the Creative Commons Attribution 4.0 International License (CC-BY 4.0).


\clearpage

\section{Limitation}
First, our current prompts and data generation procedures target Python programs. While this choice reflects the prevalence of Python in existing coding benchmarks, it may limit immediate applicability to other programming languages. That said, the underlying Orthogonal Defect Classification (ODC) categories are language-agnostic, and adapting \pdb to new languages primarily requires modifying in-context examples and language-specific sub-categories, rather than redesigning the framework.

Second, this work focuses on benchmark construction and evaluation, but does not train or post-train models using \pdb{} signals. Future work should incorporate \pdb{} into supervised fine-tuning, reinforcement learning, or self-improvement pipelines to test whether precision-aware objectives can reduce over-editing while preserving functional correctness.

Finally, although our edit-level precision and bug-level recall metrics provide a more accurate characterization of debugging behavior than unit-test–only evaluation, they may still fail to capture certain correct but semantically equivalent fixes. Incorporating more flexible semantic evaluation mechanisms, such as LLM-as-a-judge, may help address these edge cases. More broadly, reliably evaluating semantic correctness of code edits remains an open problem.


\end{document}